\documentclass[reprint,superscriptaddress,amsmath,amssymb,aps,prl,floatfix]{revtex4-2}

\usepackage{dcolumn}    
\usepackage{bm}         

\usepackage{rotating}
\usepackage[colorinlistoftodos]{todonotes}
\usepackage{amsmath,amsfonts,amsthm,bbold}

\newcommand{\be}{\begin{equation}} 
\newcommand{\ee}{\end{equation}}

\newcommand{\dif}{\scalebox{0.7}{\ensuremath{\Delta}}}

\newcommand{\n}[1]{\mathrm{#1}}
\newcommand{\cl}[1]{\mathcal{#1}}

\newcommand{\bra}[1]{\ensuremath{\langle#1|}}
\newcommand{\ket}[1]{\ensuremath{|#1\rangle}}

\newcommand{\proba}{\ensuremath{\mathbb{P}}}

\newcommand{\half}{\ensuremath{\frac{1}{2}}}

\newcommand{\bexp}{\scalebox{1.2}{\ensuremath{\n{e}}}}

\newcommand{\Gerr}{\ensuremath{\Gamma_\n{err}}}
\newcommand{\Gqec}{\ensuremath{\Gamma_\n{qec}}}
\newcommand{\Geff}{\ensuremath{\Gamma_\n{eff}}}
\newcommand{\Gest}{\ensuremath{\hat\Gamma}_\n{err}}

\newcommand{\oeff}{\ensuremath{\omega_\n{eff}}}

\newcommand{\oest}{\ensuremath{\hat{\omega}}}

\newcommand{\topt}{\ensuremath{\tau_\n{opt}}}
\newcommand{\kopt}{\ensuremath{k_\n{opt}}}

\usepackage{pdfpages}
\makeatletter
\AtBeginDocument{\let\LS@rot\@undefined}
\makeatother

\makeatletter
\renewcommand*\env@matrix[1][\arraystretch]{%
  \edef\arraystretch{#1}%
  \hskip -\arraycolsep
  \let\@ifnextchar\new@ifnextchar
  \array{*\c@MaxMatrixCols c}}
\makeatother

\usepackage{tikz}
\usetikzlibrary{quantikz}
\usepackage{adjustbox}

\usepackage{ragged2e}

\usepackage{printlen}
\usepackage{graphicx}

\usepackage[innercaption]{sidecap}
\sidecaptionvpos{figure}{c}

\usepackage{subcaption}

\DeclareCaptionJustification{justified}{\justifying}
\graphicspath{{figures/}}
\newcommand{\figref}[1]{Fig.~\ref{#1}}

\usepackage[T1]{fontenc}
\usepackage[utf8]{inputenc}
\usepackage{pgfplots}
\DeclareUnicodeCharacter{2212}{−}
\usepgfplotslibrary{groupplots,dateplot}
\usetikzlibrary{patterns,shapes.arrows}
\pgfplotsset{compat=newest}

\begin{document}
\preprint{APS/123-QED}

\title{Bias in error-corrected quantum sensing}

\author{Ivan Rojkov}
  \email{irojkov@ethz.ch}
  \affiliation{
    Institute for Quantum Electronics,
    ETH Z\"{u}rich, 8093 Z\"{u}rich, Switzerland.
    }

\author{David Layden}
  \email{Current address: IBM Quantum, Almaden Research Center, San Jose, California 95120, USA}
  \affiliation{
    Research Laboratory of Electronics and
    Department of Nuclear Science and Engineering,
    Massachusetts Institute of Technology,
    Cambridge, Massachusetts 02139, USA
    }

\author{Paola Cappellaro}
  \affiliation{
    Research Laboratory of Electronics and
    Department of Nuclear Science and Engineering,
    Massachusetts Institute of Technology,
    Cambridge, Massachusetts 02139, USA
    }

\author{Jonathan Home}
  \affiliation{
    Institute for Quantum Electronics,
    ETH Z\"{u}rich, 8093 Z\"{u}rich, Switzerland.
    }

\author{Florentin Reiter}
  \email{freiter@phys.ethz.ch}
  \affiliation{
    Institute for Quantum Electronics,
    ETH Z\"{u}rich, 8093 Z\"{u}rich, Switzerland.
    }

\date{\today}

\begin{abstract}
The sensitivity afforded by quantum sensors is limited by decoherence. Quantum error correction (QEC) can enhance sensitivity by suppressing decoherence, but it has a side-effect: it biases a sensor's output in realistic settings. If unaccounted for, this bias can systematically reduce a sensor's performance in experiment, and also give misleading values for the minimum detectable signal in theory. We analyze this effect in the experimentally-motivated setting of continuous-time QEC, showing both how one can remedy it, and how incorrect results can arise when one does not.
\end{abstract}

\maketitle


\noindent
\textit{Introduction.---}
Quantum sensors use quantum systems and effects to sense an external signal $V(t)$ in their environment, such as electromagnetic fields, temperature or pressure \cite{degen_quantum_2017}. They also, however, experience decoherence due to this same environment, which limits their sensitivity in practice. Techniques to suppress decoherence, without equally suppressing the signal, are therefore of central importance in quantum sensing \cite{viola_dynamical_1998,taylor_high-sensitivity_2008,cai_robust_2012,lang_dynamical-decoupling-based_2015,lazariev_dynamical_2017,schmitt_submillihertz_2017,genov_mixed_2019,ban_photon-echo_1998}. Quantum error correction (QEC) is currently emerging as an important technique to this end, and has attracted substantial theoretical and experimental interest of late \cite{ozeri_heisenberg_2013,kessler_quantum_2014,dur_improved_2014,arrad_increasing_2014,herrera-marti_quantum_2015,plenio_sensing_2016,gefen_parameter_2016,reiter_dissipative_2017,matsuzaki_magnetic-field_2017, unden_quantum_2016,cohen_demonstration_2016, bergmann:2016, sekatski:2017, zhou:2018, demkowicz:2017, layden:2018, layden:2019, gorecki:2020}. Indeed, while QEC was first developed in the context of quantum computing, the substantially less demanding experimental requirements of quantum sensing make the latter application particularly attractive for QEC in the near term \cite{lidar:2013}. It is therefore essential to address the unique challenges that arise from using QEC for sensing, which often lack close analogues in quantum computing (or quantum communication).

Quantum sensing generally works by preparing a sensor in a known initial state, letting it evolve under dynamics that depend on the signal $V(t)$, and then measuring the sensor's final state so as to estimate the signal. In error-corrected quantum sensing, the initial state is a logical state of a QEC code, and errors are repeatedly corrected while the signal imprints on the sensor, until the latter is finally read out. The error correction can either be performed at discrete times through measurement and feedback, or continuously by engineering appropriate dissipative terms in the sensor's dynamics. Most analyses of QEC for sensing to date have focused on the leading-order effective dynamics under very frequent/strong error correction (in the discrete/continuous pictures respectively) to show that QEC can---in principle---completely suppress decoherence in a sensor. 

In this Letter we instead examine quantum sensing under QEC of manifestly finite frequency/strength. As expected, we find that such QEC reduces, but does not fully eliminate decoherence in a sensor. Crucially, however, we also find that it can systematically bias a sensor's output, i.e., introduce inaccuracy that is not reduced by averaging over many runs. Using elementary assumptions, we determine under which condition this bias arises and illustrate it with a canonical example of error-corrected quantum sensing. Fundamentally, it appears from a nonzero delay between an error and its subsequent correction, during which time a sensor typically evolves differently than if no error had occurred. This effect would clearly reduce a sensor's performance in experiment if one failed to account for it. Similarly, we use biased estimator theory to show that a standard analysis can give misleading values of sensitivity in theory. Finally, we give a simple method to remedy both issues through post-processing of measurement data.

\begin{figure*}[ht!]
  \centering
  \includegraphics[width=\linewidth]{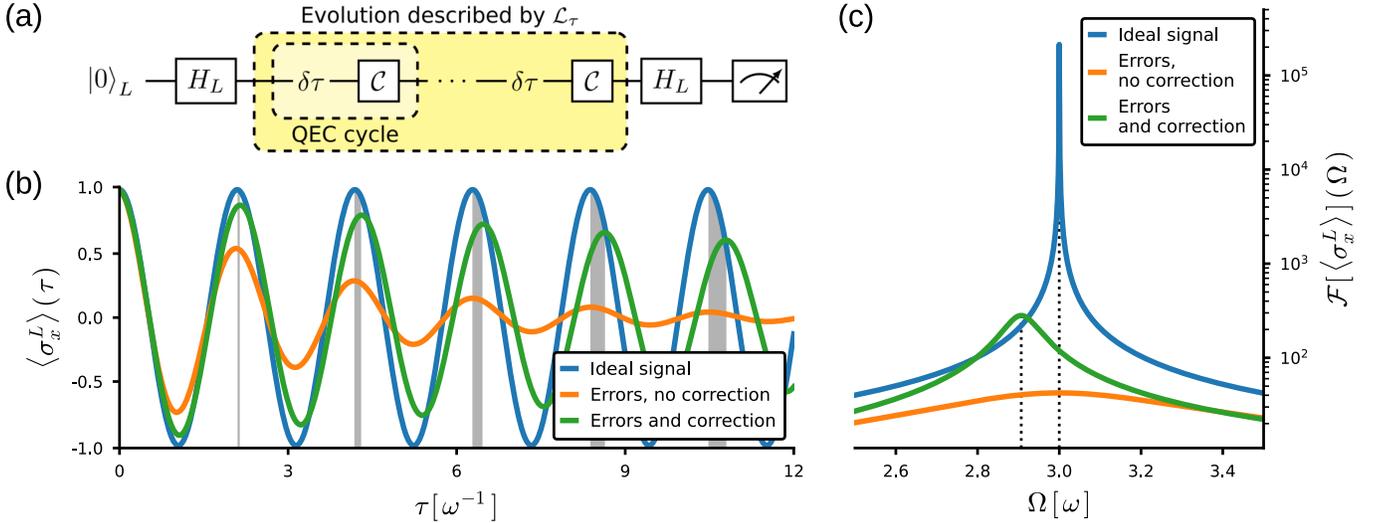}
  {\phantomsubcaption\label{fig:ramsey_circuit}}
  {\phantomsubcaption\label{fig:ramsey_parity}}
  {\phantomsubcaption\label{fig:ramsey_fourier}}
  \captionsetup{justification=justified}
  \caption{Bias in error-corrected quantum sensing. (a) Error-corrected Ramsey sequence represented as a quantum circuit where $H_L$ are logical Hadamard gates, $\cl{C}$ represents detection and correction processes and $\delta\tau$ free evolution intervals under the signal with frequency $\omega$. The overall dynamics during the sensing time $\tau$ is described by $\cl{L}_\tau$. In (b), we plot the expectation value $\langle \sigma_x^L \rangle$ as a function of the free evolution time $\tau$ for three different sensing models: ideal, uncorrected and error-corrected. The parameters used for the simulation are $\Gerr=0.1\omega$, $\Gqec=5\omega$. Panel (c) shows the Fourier transform $\mathcal{F}$ of the functions from \figref{fig:ramsey_parity}.}
  \label{fig:Fig1}
\end{figure*}


\medskip \noindent
\textit{Setting.---} We consider the problem of estimating a DC signal $V(t)=V$ using quantum sensor comprising $n$ identical qubits with a Hamiltonian
\be \label{eq:hamiltonian}
    H = \sum_{j=1}^n
    \left( 
    \frac{\omega_q}{2}\,\sigma_z^{(j)}  + \xi \,V\,\sigma_z^{(j)} \right)
      = \sum_{j=1}^n \frac{\omega}{2}\,\sigma_z^{(j)},
\ee
where $\omega_q$ is the qubits' zero-field splitting and $\xi$ is their coupling strength to the environment ($\hbar=1$). We define for convenience the combined energy gap ${\omega := \omega_q + 2\,\xi\,V}$. We also suppose that the sensor undergoes decoherence described by a Lindblad equation. Following Refs.~\cite{ozeri_heisenberg_2013,dur_improved_2014,arrad_increasing_2014,cohen_demonstration_2016,reiter_dissipative_2017}, we assume this decoherence to be primarily due to bit flips affecting each qubit independently at the same rate $\Gamma_\text{err}$, described by the Lindblad jump operators $L^{(j)}_\n{err}=\sqrt{\Gerr}\,\sigma_x^{(j)}$~\footnote{Instead of considering $H\propto\sigma_z$ and $L_\n{err}\propto\sigma_x$, some works looked at the problem from the opposite point of view by taking $H\propto\sigma_x$ and $L_\n{err}\propto\sigma_z$. This only changes the codewords used by the sensing protocol but not the conclusions of our work since the signal and the noise remain perpendicular in both cases.}. Note that this is a pessimistic model in the sense that dynamical decoupling, one of the main techniques for suppressing decoherence in quantum sensors, is largely incompatible with both DC signals and Lindbladian decoherence \cite{degen_quantum_2017}.

This decoherence limits the sensor's achievable sensitivity. It can be suppressed, and the sensitivity enhanced, by using the bit-flip code where $\ket{0}_L = \ket{0\dots0}$ and $\ket{1}_L = \ket{1 \dots 1}$ for $n \ge 3$ qubits. The idea is to mimic a Ramsey interferometry protocol \cite{ramsey_molecular_1950} at the logical level. One first prepares a state $\frac{1}{\sqrt{2}}(\ket{0}_L+\ket{1}_L)$ (say by applying a logical Hadamard, $H_L$, to $\ket{0}_L$),
then lets the sensor evolve for a time $\tau$ all the while detecting and correcting $L_\text{err}^{(j)}$ errors. One then applies $H_L$ to the time-evolved state and measures in the $\{\ket{0}_L, \ket{1}_L\}$ basis, effectively measuring $\sigma_x^L = \ket{0}\!\bra{1}_L + \ket{1} \! \bra{0}_L$.

Broadly speaking, errors can be detected and corrected in two ways, both illustrated in \figref{fig:ramsey_circuit}. One is to periodically measure the parity between qubits at intervals $\delta \tau$ to infer whether an error has occurred, and then to apply appropriate feedback. The other is to engineer additional jump operators in the sensor's Liouvillian $\mathcal{L}_\tau$ that implement a continuous analogue of the former scheme. We focus here on the latter, continuous type for mathematical convenience. For $n=3$, the QEC jump operators have the form \cite{reiter_dissipative_2017}
\be \label{eq:jump_qec}
	L_\n{qec}^{(j)} = \sqrt{\Gqec}\,\sigma_x^{(j)}\,\frac{1-\sigma_z^{(j)}\,\sigma_z^{(k)}}{2}\,\,\frac{1-\sigma_z^{(j)}\,\sigma_z^{(l)}}{2}
\ee
for $\{j,k,l\} = \{1,\,2,\,3\}$. Notice that $L_\text{qec}^{(j)}$ flips qubit $j$ at a rate $\Gamma_\text{qec}$ if and only if it disagrees with the other two qubits. The sensor's overall dynamics is then given by the Lindblad equation $\dot{\rho} = \mathcal{L}_\tau(\rho)$ with the Liouvillian
\be \label{eq:master_eq}
	\mathcal{L}_\tau(\rho) = -i\,[ H , \, \rho ] + \sum_j \cl{D}[\,L_\n{err}^{(j)}\,](\rho) + \sum_j \cl{D}[\,L_\n{qec}^{(j)}\,](\rho)
\ee
where 
\[
    \cl{D}[\,L_k^{(j)}\,](\rho)\coloneqq  L_k^{(j)}\,\rho\,L_k^{(j)\,\dagger} - \frac{1}{2}\left(L_k^{(j)\,\dagger} L_k^{(j)}\,\rho\, + \,\rho\, L_k^{(j)\,\dagger} L_k^{(j)}\right)
\]
for a jump operator $L_k^{(j)}$.  The bias we are concerned with arises from both types QEC, with $\Gqec$ and $\delta\tau^{-1}$ playing the same role.

\medskip \noindent
\textit{Imperfect error-corrected sensing.---}Repeating the protocol in \figref{fig:ramsey_circuit} gives a set of binary measurement results $\{X_\tau\}$, where $X_\tau = \pm1$ for $\ket{0}_L$/$\ket{1}_L$ outcomes respectively. One can then infer $\omega$ with the least-squares estimator
\be \label{eq:estimator}
  \oest=\underset{\omega}{\arg\min} \sum_\tau \Big[\,X_\tau - \langle \sigma_x^L  (\tau,\omega,\Gerr,\Gqec) \rangle\,\Big]^2,
\ee
which minimizes the discrepancy between the observed and expected results. For a single, noiseless qubit ${\langle \sigma_x \rangle = \cos(\omega \tau)}$. Similarly, the three-qubit repetition code gives $\langle \sigma_x^L \rangle = \cos(3\omega \tau)$ in the limit of infinitely frequent/strong QEC ($\Gamma_\text{qec}$ or $\delta \tau^{-1} \rightarrow \infty$)---producing, in effect, a noiseless qubit with a stronger dependence on $\omega$ at the logical level. With finite frequency/strength QEC, the logical qubit will gradually decohere, albeit hopefully more slowly than the physical qubits. One might expect that in this latter, realistic setting, $\langle \sigma_x^L \rangle$ would behave like in the ideal case, but with an additional exponential decay that vanishes as $\Gamma_\text{err} \rightarrow 0$:
\begin{equation}
\label{eq:conjecture}
\langle \sigma_x^L \rangle
\stackrel{?}{=}
e^{-3 \Gamma_\text{err} \tau} \cos(3 \omega \tau).
\end{equation}
We will show, however, that while the true expression for $\langle \sigma_x^L \rangle$ indeed has this form, the coefficients must be replaced by distinct, effective ones $\Gamma_\text{err} \rightarrow \Gamma_\text{eff}$ and ${\omega \rightarrow \omega_\text{eff}}$ given by Eq.~\eqref{eq:eff_coeffs}.

Finding $\langle \sigma_x^L \rangle$ amounts to solving a system of first order differential equations describing evolution under $\mathcal{L}_\tau$. We sketch the process here, relegating the details to the Supplemental Material~\footnote{See Supplemental Material.}. To get a tractable system, we make the approximation that rare, weight-2 errors cause the state to vanish. This quickly leads to the solution $\langle \sigma_x^L \rangle=2 \text{Re}(q)$, where
\be \label{eq:my_sol}
  \begin{split}
q(\tau) &= \,
		C_+\,\bexp^{\,\half\tau \left(-\Gqec - 6\Gerr - 4 i \omega + \sqrt{D}\right)} \\ 
		&- C_-\,\bexp^{\,\half\tau \left(-\Gqec - 6 \Gerr - 4 i \omega - \sqrt{D}\right)}.
	\end{split}
\ee
Here $C_\pm$ are normalization constants (discussed below and in \cite{Note2}) and $D=\Gqec^2+12\,\Gqec\,\Gerr+12\,\Gerr^2-4i\,\Gqec\,\omega-4\,\omega^2$ is a discriminant. Expanding this expression for a large---but finite---QEC rate then gives
\be
  \begin{split}
    &\langle \sigma_x^L \rangle = 
        2 C_+\,\bexp^{-6\,\frac{\Gerr}{\Gqec}\,\Gerr\,\tau}\,
        \cos\left[3\,\omega(1 - 2\,\frac{\Gerr}{\Gqec})\tau\right] \\[6pt]
      &-2C_-\,\bexp^{-(\Gqec +6\,\Gerr -6\,\frac{\Gerr^2}{\Gqec})\tau}\,
        \cos\left[\omega(1 + 6\,\frac{\Gerr}{\Gqec})\tau\right] 
  \end{split}
  \label{eq:my_sol_simplified}
\ee
where $C_+ = \half - \frac{3}{2}\frac{\Gerr}{\Gqec}$ and $C_- = - \frac{3}{2}\frac{\Gerr}{\Gqec}$. Simplifying further still, we arrive at 
\be
\label{eq:approximate_XL}
  \langle \sigma_x^L \rangle \approx
  \n{e}^{-3\,\Geff\,\tau\, }\cos[\,3\,\oeff\,\tau\,]
\ee
[cf.\ \eqref{eq:conjecture}] where
\be
    \begin{split}
    \label{eq:eff_coeffs}
    \Geff = 2\,\frac{\Gerr}{\Gqec}\Gerr \qquad \quad
    \oeff = \omega\,\left(1-2\,\frac{\Gerr}{\Gqec}\right)
    \end{split}
\ee
to first order in $\Gamma_\text{qec}^{-1}$, as claimed above. In other words, imperfect QEC not only leads to logical decoherence; it also shifts the oscillation frequency of $\langle \sigma_x^L \rangle$. This is because the sensor acquires a relative phase at a rate $3\omega$ in the absence of errors, but only at a rate $\omega$ in the brief intervals between an error and its subsequent correction. Therefore errors reduce the average precession rate $\omega_\text{eff}$, and so $\omega_\text{eff} < \omega$ for finite $\Gamma_\text{qec}$. This bias is illustrated in the time and frequency domains in Figs.~\ref{fig:ramsey_parity} and  \ref{fig:ramsey_fourier} respectively. 

These figures show additionally that the bias is more pronounced in the error-corrected sensing situation than in an uncorrected one. This comes primarily from the fact that the leading-order behaviour of the frequency scaling factor is in the latter case quadratic in system's parameters~\cite{Note2} and not linear as shown in Eq.~\eqref{eq:eff_coeffs}. The finite QEC rate will then act as an amplifier of the bias which in the non-corrected case was likely undetectable. While the fact that the transition from $\Gqec=0$ to $\Gqec>0$ does not fully eliminate the original bias is unsurprising, its transformation from a quadratic to a linear function in the dominant limits is an unexpected result.

\medskip \noindent
\textit{Biasedness condition.---}
As mentioned previously, an alternative pragmatic representation of the qubits' evolution during the error-corrected sensing period is the repeated application of detection and correction processes $\cl{C}$ interspersed with sensing time intervals $\delta\tau$.

The evolution of the generic sensor can then be described by a set of three quantum channels: $\cl{N}$, $\cl{U}_{\delta\tau}$ and $\cl{C}$. The first element represents noise mechanisms which can be methodically approximated by a Pauli channel~\cite{wallman_noise_2016,harper_efficient_2020,ware_experimental_2021}. It is convenient to split it into noiseless and erroneous parts $\cl{N}=p_\cl{I}\,\cl{I}+p_\cl{N}\cl{N}'$ where $p_\cl{I} = 1 - p_\cl{N}$ is the probability that the state stays unharmed. $\cl{U}_{\delta\tau}(\rho)=U(\delta\tau)\,\rho\, U(\delta\tau)^\dagger$ with ${U(\delta\tau)=\exp\left(-i H\,\delta\tau \right)}$ expresses the unitary free evolution in the interval $\delta\tau$. The overall dynamics during the sensing time $\tau$ can then be approximated by the iteration of these three channels $c$\,-\,times, such that $c\,\delta\tau=\tau$,
\be \label{eq:disc_evolution}
\begin{split}
  \rho \longmapsto &\left[\,\prod_{k=1}^{c} \cl{C} \circ \,\cl{U}_{\delta\tau} \circ \cl{N} \right]\!(\rho)\,\,= \\
  &=\Bigl[\,p_\cl{I}\,\cl{C}\circ\,\cl{U}_{\delta\tau} + p_\cl{N}\,\cl{C} \circ \,\cl{U}_{\delta\tau} \circ \cl{N}' \Bigr]^c(\rho)
\end{split}
\ee
This expression can be simplified using the binomial theorem which is permitted here since in the logical subspace $\cl{C}\circ\,\cl{U}_{\delta\tau}$ commutes with $\cl{C}\circ\,\cl{U}_{\delta\tau}\circ \cl{N}'$~\cite{Note2}. Finally, the discrete evolution can be reduced to the following form
\be
\label{eq:disc_evolution_binomial}
    \rho \longmapsto \sum_{k=0}^{c} \begin{pmatrix} c \\ k \end{pmatrix} p_\cl{I}^{c-k}\,p_\cl{N}^k \Bigl[\cl{C} \circ \,\cl{U}_{\delta\tau} \circ \cl{N}'\circ\,\cl{U}_{\delta\tau}^{-1} \Bigr]^{k}\,(\rho_{\tau})
\ee
with $\rho_{\tau}$ being the density matrix in the case of an ideal sensing protocol of duration $\tau$, i.e. $\rho_{\tau}=\cl{U}_{\tau}(\rho)=\cl{U}_{\delta\tau}^{c}(\rho)$. Eq.~\eqref{eq:disc_evolution_binomial} reveals that finite rate QEC will systematically reduce the sensor's performance unless the erroneous subspaces evolve in the exactly same manner as the logical one. In this scenario, $\cl{U}_{\delta\tau}\circ\cl{N'}$ and $\cl{U}_{\delta\tau}^{-1}$ would commute and reduce the expression to a unit channel. This is for instance the case in continuous-variable encodings where due to the continuity between the logical and error subspaces their evolutions do not differ. An example of electric field sensing using such encodings is discussed in the Supplemental Material~\cite{Note2}. The commutation relation $\bigl[\,\cl{U}_{\delta\tau}\circ\cl{N'},\,\cl{U}_{\delta\tau}^{-1}\,\bigr](\rho_L)=0$ offers thus any experimenter a sanity check procedure for the biasedness of their error corrected sensing protocol. 

Furthermore Eq.~\eqref{eq:disc_evolution_binomial} represents an expected value of a CPTP map function $\bigl[\cl{C} \circ \,\cl{U}_{\delta\tau} \circ \cl{N}'\circ\,\cl{U}_{\delta\tau}^{-1} \bigr]^{X}$ with $X$ being a random variable following a binomial distribution. It can thus be approximated by a Taylor expansion around $\mathbb{E}[X]=c\,p_\cl{N}$,
\be \label{eq:disc_evolution_approx}
  \rho \rightsquigarrow \Bigl[\cl{C} \circ \,\cl{U}_{\delta\tau} \circ \cl{N}'\circ\,\cl{U}_{\delta\tau}^{-1} \Bigr]^{c\,p_\cl{N}} \,(\rho_{\tau})
\ee
which is correct up to the second order in the expansion. This constitutes the biased evolution of the system without the exponential decoherence which appears from higher-order terms and central moments of ${X\sim \rm{Bin}(c,p_\cl{N})}$~\cite{Note2}. In conclusion, to determine the amount of bias that a finite-rate QEC introduces in the outcome of a sensing protocol one must evaluate the expected value of the measured operator given in Eq.~\eqref{eq:disc_evolution_approx}. In the Supplemental Material, we perform the analysis of our error-corrected Ramsey setting using this procedure and verify its agreement with simulations.

\medskip \noindent
\textit{Biased parameter estimation.---}
Another method to assess the presence of a bias in a sensing protocol is to verify some statistical inequalities. A direct application of Eq.~\eqref{eq:conjecture}, i.e., equating $\omega$ with $\omega_\text{eff}$, violates a fundamental bound of estimation theory known as Cram\'{e}r-Rao lower bound~\cite{cramer_mathematical_1999}. On the contrary, the simple remedy in \eqref{eq:eff_coeffs} almost completely resolves the issue, despite arising from a perturbative expansion. This result~\cite{Note2} obatained using Monte-Carlo simulations of the presented system shows that the na\"ive estimator arising from Eq.~\eqref{eq:conjecture} is biased. Perhaps more perniciously, we also show equating $\omega$ and $\omega_\text{eff}$ can give misleadingly optimistic figures of merit for error-corrected quantum sensors.


\medskip \noindent
\textit{Minimum detectable signal.---} 
A central figure of merit for a quantum sensor is the minimum detectable signal $|\delta \omega|$, defined as the smallest value of $\omega$ giving a unit signal to noise ratio \cite{degen_quantum_2017}. Here $|\delta \omega| = 1/\sqrt{N \mathcal{I}(\omega)}$. The ideal case of $\langle \sigma_x^L \rangle = \cos(3 \omega \tau)$ of perfect QEC, or noiseless qubits, gives a lower bound of $|\delta \omega|(\tau) \ge (9\,N\,\tau^2)^{-1/2}$ called the Standard Quantum Limit (SQL). 

A straightforward application of Eq.~\eqref{eq:conjecture} would suggest a minimum detectable signal of
\be \label{eq:sensitivity}
  |\delta\omega|(\tau)
  \stackrel{?}{=}
  \sqrt{\frac{1-\bexp^{-6\,\Gerr\,\tau}\,\cos^2[\,3\,\omega\,\tau\,]}
  {N\,9\,\tau^2\,\bexp^{-6\,\Gerr\,\tau}\,\sin^2[\,3\,\omega\,\tau\,]}} .
\ee
This expression equals the frequency uncertainty derived by \citet{huelga_improvement_1997}.
It is minimized (smaller $|\delta \omega|$ is better) at the optimal sensing time
\be \label{eq:tau_k_opt}
  \topt = \frac{\pi}{2}\frac{\kopt}{3\,\omega}
  \quad\,\,\textnormal{with}\,\,\quad
  \kopt =  \left\lfloor\,\frac{2}{\pi} \frac{\omega}{\Gerr}\,\right\rceil,
\ee
where $\lfloor\,\cdot\,\rceil$ indicates the rounding to the nearest integer. In contrast, the corresponding expression from Eq.~\eqref{eq:approximate_XL} (i.e., accounting for $\omega \neq \omega_\text{eff}$ and $\Gamma_\text{err} \neq \Gamma_\text{eff}$) is
\be \label{eq:my_sensitivity}
  |\delta\omega|(\tau)=
  \sqrt{\frac{1-\bexp^{-6\,\Geff\,\tau}\,\cos^2[\,3\,\oeff\,\tau\,]}
  {N\,9\,\tau^2(\partial_\omega\oeff)^2\,\bexp^{-6\,\Geff\,\tau}\,\sin^2[\,3\,\oeff\,\tau\,]}}
\ee
where $\partial_\omega\oeff$ comes from the derivative of $\langle \sigma_x^L\rangle(\tau)$ with respect to $\omega$. (The expression for $|\delta \omega|$ from the full Eq.~\eqref{eq:my_sol} is lengthy and can be found in \cite{Note2}.) Since the former is equal to $1-2\frac{\Gerr}{\Gqec}$, independent of $\tau$, the optimal sensing time is of the same form as in Eq.~\eqref{eq:tau_k_opt} but with the effective frequency and error rate substituted. In short, failing to account for the difference between $(\omega, \Gamma_\text{err})$ and $(\omega_\text{eff},\Gamma_\text{eff})$ would suggest an overly optimistic value of $|\delta \omega|$, off by a factor of $\sim\!\left|1-2\frac{\Gerr}{\Gqec}\right|$. We graphically illustrate this inconsistency as follows:

\begin{figure}[t]
  \centering
  \includegraphics[width=\linewidth]{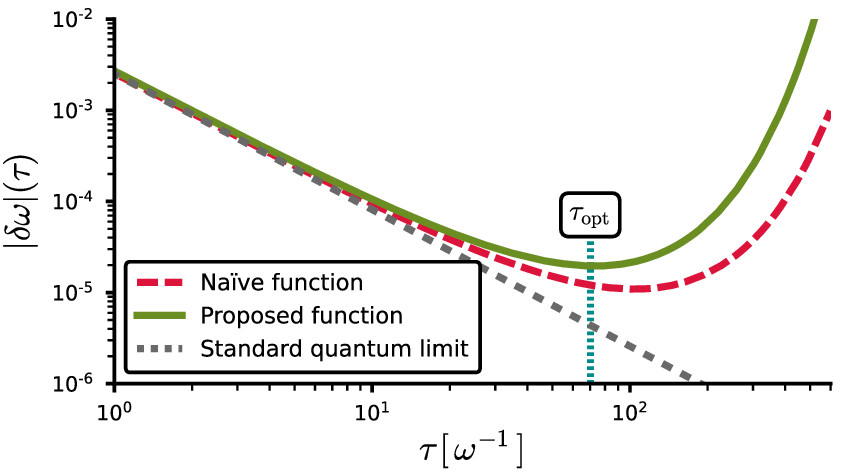}
  \caption{Minimum detectable signal as a function of the sensing time $\tau$ for two different functions: na\"ive (dashed) and proposed (solid) using Eq.~\eqref{eq:sensitivity} and \eqref{eq:my_sensitivity}, respectively. Both functions were plotted with the same values of $\oest$ and $\Gest$, for $\Gerr=0.2\omega$ and $\Gqec=16.6\omega$. The optimal sensing time $\topt$ calculated with Eq.~\eqref{eq:tau_k_opt} coincides with the minimum of the proposed function, thus establishing it as the correct model.}
  \label{fig:Fig3}
\end{figure}

We consider a set of observations for which a frequency $\oest$ and an error rate $\Gest$ were estimated using the unadjusted expression in Eq.~\eqref{eq:conjecture}. Then, in the na\"ive scenario with $\oest\equiv\omega$ and $\Gest\equiv\Gerr$, we insert them in Eq.~\eqref{eq:sensitivity} and calculate a first curve. In our proposed scenario, where we are aware of the existence of a bias, the estimated parameters stand for $\oest\equiv\oeff$ and $\Gest\equiv\Geff$. Substituting them into Eq.~\eqref{eq:my_sensitivity} yields a second curve. These two functions are presented in \figref{fig:Fig3} as dashed and solid lines, respectively. Although both of them satisfy the standard quantum limit, only one can correctly reflect the true minimum detectable signal. We utilize the optimal sensing time to identify the correct function. To this end, we calculate $\topt$ using Eq.~\eqref{eq:tau_k_opt} for both scenarios and plot the result in \figref{fig:Fig3}. As we can see, $\topt$ comes out the same for both scenarios. This can be understood from the fact that it can either way be written  as $\topt\approx(\Gest)^{-1}$. Yet, the computed $\topt$ coincides with the minimum of the proposed function, while deviating from the na\"ive one. We thus conclude that only the proposed model delivers the true result for the minimum detectable signal.


\medskip \noindent
\textit{Discussion and conclusion.---}
In summary, we demonstrate in this work that quantum error correction of a finite frequency/strength in a quantum sensor not only introduces logical decoherence, but can also produce a systematic bias in the sensor's output. We showed that this effect can lead to overly optimistic---and even nonsensical---parameter estimates if unaccounted for, and proposed a remedy through post-processing. An interesting open question is whether it is possible to design QEC codes for sensing that do not introduce such a bias in the first place.

While we have focused on the bit-flip code for concreteness, we expect the phenomenon discussed here to arise much more generally when the error operators do not commute with the sensor's Hamiltonian. Similarly, our results are not limited to Ramsey sensing schemes, even though we have focused on this approach. This work shows that nontrivial, and perhaps unexpected, effects can arise when performing QEC beyond the idealized limits considered in early proposals. We expect that much remains to be discovered in this direction on the way to realizing useful QEC.

\textit{Acknowledgments.---} I.R. thanks D.~Roschewitz for helpful comments on the manuscript. This work was supported by the Swiss National Science Foundation (SNSF) through the National Centre of Competence in Research - Quantum Science and Technology (NCCR QSIT) grant 51NF40–160591. I.R. and F.R. acknowledge financial support by the Swiss National Science Foundation (Ambizione grant no. PZ00P2$\_$186040). 


\bibliography{references}

\begin{thebibliography}{54}%
\makeatletter
\providecommand \@ifxundefined [1]{%
 \@ifx{#1\undefined}
}%
\providecommand \@ifnum [1]{%
 \ifnum #1\expandafter \@firstoftwo
 \else \expandafter \@secondoftwo
 \fi
}%
\providecommand \@ifx [1]{%
 \ifx #1\expandafter \@firstoftwo
 \else \expandafter \@secondoftwo
 \fi
}%
\providecommand \natexlab [1]{#1}%
\providecommand \enquote  [1]{``#1''}%
\providecommand \bibnamefont  [1]{#1}%
\providecommand \bibfnamefont [1]{#1}%
\providecommand \citenamefont [1]{#1}%
\providecommand \href@noop [0]{\@secondoftwo}%
\providecommand \href [0]{\begingroup \@sanitize@url \@href}%
\providecommand \@href[1]{\@@startlink{#1}\@@href}%
\providecommand \@@href[1]{\endgroup#1\@@endlink}%
\providecommand \@sanitize@url [0]{\catcode `\\12\catcode `\$12\catcode
  `\&12\catcode `\#12\catcode `\^12\catcode `\_12\catcode `\%12\relax}%
\providecommand \@@startlink[1]{}%
\providecommand \@@endlink[0]{}%
\providecommand \url  [0]{\begingroup\@sanitize@url \@url }%
\providecommand \@url [1]{\endgroup\@href {#1}{\urlprefix }}%
\providecommand \urlprefix  [0]{URL }%
\providecommand \Eprint [0]{\href }%
\providecommand \doibase [0]{https://doi.org/}%
\providecommand \selectlanguage [0]{\@gobble}%
\providecommand \bibinfo  [0]{\@secondoftwo}%
\providecommand \bibfield  [0]{\@secondoftwo}%
\providecommand \translation [1]{[#1]}%
\providecommand \BibitemOpen [0]{}%
\providecommand \bibitemStop [0]{}%
\providecommand \bibitemNoStop [0]{.\EOS\space}%
\providecommand \EOS [0]{\spacefactor3000\relax}%
\providecommand \BibitemShut  [1]{\csname bibitem#1\endcsname}%
\let\auto@bib@innerbib\@empty
\bibitem [{\citenamefont {Degen}\ \emph {et~al.}(2017)\citenamefont {Degen},
  \citenamefont {Reinhard},\ and\ \citenamefont
  {Cappellaro}}]{degen_quantum_2017}%
  \BibitemOpen
  \bibfield  {author} {\bibinfo {author} {\bibfnamefont {C.~L.}\ \bibnamefont
  {Degen}}, \bibinfo {author} {\bibfnamefont {F.}~\bibnamefont {Reinhard}},\
  and\ \bibinfo {author} {\bibfnamefont {P.}~\bibnamefont {Cappellaro}},\
  }\bibfield  {title} {\bibinfo {title} {Quantum sensing},\ }\href
  {https://doi.org/10.1103/RevModPhys.89.035002} {\bibfield  {journal}
  {\bibinfo  {journal} {Rev. Mod. Phys.}\ }\textbf {\bibinfo {volume} {89}},\
  \bibinfo {pages} {035002} (\bibinfo {year} {2017})}\BibitemShut {NoStop}%
\bibitem [{\citenamefont {Viola}\ and\ \citenamefont
  {Lloyd}(1998)}]{viola_dynamical_1998}%
  \BibitemOpen
  \bibfield  {author} {\bibinfo {author} {\bibfnamefont {L.}~\bibnamefont
  {Viola}}\ and\ \bibinfo {author} {\bibfnamefont {S.}~\bibnamefont {Lloyd}},\
  }\bibfield  {title} {\bibinfo {title} {Dynamical suppression of decoherence
  in two-state quantum systems},\ }\href
  {https://doi.org/10.1103/PhysRevA.58.2733} {\bibfield  {journal} {\bibinfo
  {journal} {Phys. Rev. A}\ }\textbf {\bibinfo {volume} {58}},\ \bibinfo
  {pages} {2733} (\bibinfo {year} {1998})}\BibitemShut {NoStop}%
\bibitem [{\citenamefont {Taylor}\ \emph {et~al.}(2008)\citenamefont {Taylor},
  \citenamefont {Cappellaro}, \citenamefont {Childress}, \citenamefont {Jiang},
  \citenamefont {Budker}, \citenamefont {Hemmer}, \citenamefont {Yacoby},
  \citenamefont {Walsworth},\ and\ \citenamefont
  {Lukin}}]{taylor_high-sensitivity_2008}%
  \BibitemOpen
  \bibfield  {author} {\bibinfo {author} {\bibfnamefont {J.~M.}\ \bibnamefont
  {Taylor}}, \bibinfo {author} {\bibfnamefont {P.}~\bibnamefont {Cappellaro}},
  \bibinfo {author} {\bibfnamefont {L.}~\bibnamefont {Childress}}, \bibinfo
  {author} {\bibfnamefont {L.}~\bibnamefont {Jiang}}, \bibinfo {author}
  {\bibfnamefont {D.}~\bibnamefont {Budker}}, \bibinfo {author} {\bibfnamefont
  {P.~R.}\ \bibnamefont {Hemmer}}, \bibinfo {author} {\bibfnamefont
  {A.}~\bibnamefont {Yacoby}}, \bibinfo {author} {\bibfnamefont
  {R.}~\bibnamefont {Walsworth}},\ and\ \bibinfo {author} {\bibfnamefont
  {M.~D.}\ \bibnamefont {Lukin}},\ }\bibfield  {title} {\bibinfo {title}
  {High-sensitivity diamond magnetometer with nanoscale resolution},\ }\href
  {https://doi.org/10.1038/nphys1075} {\bibfield  {journal} {\bibinfo
  {journal} {Nat. Phys.}\ }\textbf {\bibinfo {volume} {4}},\ \bibinfo {pages}
  {810} (\bibinfo {year} {2008})}\BibitemShut {NoStop}%
\bibitem [{\citenamefont {Cai}\ \emph {et~al.}(2012)\citenamefont {Cai},
  \citenamefont {Naydenov}, \citenamefont {Pfeiffer}, \citenamefont
  {McGuinness}, \citenamefont {Jahnke}, \citenamefont {Jelezko}, \citenamefont
  {Plenio},\ and\ \citenamefont {Retzker}}]{cai_robust_2012}%
  \BibitemOpen
  \bibfield  {author} {\bibinfo {author} {\bibfnamefont {J.-M.}\ \bibnamefont
  {Cai}}, \bibinfo {author} {\bibfnamefont {B.}~\bibnamefont {Naydenov}},
  \bibinfo {author} {\bibfnamefont {R.}~\bibnamefont {Pfeiffer}}, \bibinfo
  {author} {\bibfnamefont {L.~P.}\ \bibnamefont {McGuinness}}, \bibinfo
  {author} {\bibfnamefont {K.~D.}\ \bibnamefont {Jahnke}}, \bibinfo {author}
  {\bibfnamefont {F.}~\bibnamefont {Jelezko}}, \bibinfo {author} {\bibfnamefont
  {M.~B.}\ \bibnamefont {Plenio}},\ and\ \bibinfo {author} {\bibfnamefont
  {A.}~\bibnamefont {Retzker}},\ }\bibfield  {title} {\bibinfo {title} {Robust
  dynamical decoupling with concatenated continuous driving},\ }\href
  {https://doi.org/10.1088/1367-2630/14/11/113023} {\bibfield  {journal}
  {\bibinfo  {journal} {New J. Phys.}\ }\textbf {\bibinfo {volume} {14}},\
  \bibinfo {pages} {113023} (\bibinfo {year} {2012})}\BibitemShut {NoStop}%
\bibitem [{\citenamefont {Lang}\ \emph {et~al.}(2015)\citenamefont {Lang},
  \citenamefont {Liu},\ and\ \citenamefont
  {Monteiro}}]{lang_dynamical-decoupling-based_2015}%
  \BibitemOpen
  \bibfield  {author} {\bibinfo {author} {\bibfnamefont {J.~E.}\ \bibnamefont
  {Lang}}, \bibinfo {author} {\bibfnamefont {R.-B.}\ \bibnamefont {Liu}},\ and\
  \bibinfo {author} {\bibfnamefont {T.~S.}\ \bibnamefont {Monteiro}},\
  }\bibfield  {title} {\bibinfo {title} {Dynamical-{Decoupling}-{Based}
  {Quantum} {Sensing}: {Floquet} {Spectroscopy}},\ }\href
  {https://doi.org/10.1103/PhysRevX.5.041016} {\bibfield  {journal} {\bibinfo
  {journal} {Phys. Rev. X}\ }\textbf {\bibinfo {volume} {5}},\ \bibinfo {pages}
  {041016} (\bibinfo {year} {2015})}\BibitemShut {NoStop}%
\bibitem [{\citenamefont {Lazariev}\ \emph {et~al.}(2017)\citenamefont
  {Lazariev}, \citenamefont {Arroyo-Camejo}, \citenamefont {Rahane},
  \citenamefont {Kavatamane},\ and\ \citenamefont
  {Balasubramanian}}]{lazariev_dynamical_2017}%
  \BibitemOpen
  \bibfield  {author} {\bibinfo {author} {\bibfnamefont {A.}~\bibnamefont
  {Lazariev}}, \bibinfo {author} {\bibfnamefont {S.}~\bibnamefont
  {Arroyo-Camejo}}, \bibinfo {author} {\bibfnamefont {G.}~\bibnamefont
  {Rahane}}, \bibinfo {author} {\bibfnamefont {V.~K.}\ \bibnamefont
  {Kavatamane}},\ and\ \bibinfo {author} {\bibfnamefont {G.}~\bibnamefont
  {Balasubramanian}},\ }\bibfield  {title} {\bibinfo {title} {Dynamical
  sensitivity control of a single-spin quantum sensor},\ }\href
  {https://doi.org/10.1038/s41598-017-05387-w} {\bibfield  {journal} {\bibinfo
  {journal} {Sci. Rep.}\ }\textbf {\bibinfo {volume} {7}},\ \bibinfo {pages}
  {6586} (\bibinfo {year} {2017})}\BibitemShut {NoStop}%
\bibitem [{\citenamefont {Schmitt}\ \emph {et~al.}(2017)\citenamefont
  {Schmitt}, \citenamefont {Gefen}, \citenamefont {Stürner}, \citenamefont
  {Unden}, \citenamefont {Wolff}, \citenamefont {Müller}, \citenamefont
  {Scheuer}, \citenamefont {Naydenov}, \citenamefont {Markham}, \citenamefont
  {Pezzagna}, \citenamefont {Meijer}, \citenamefont {Schwarz}, \citenamefont
  {Plenio}, \citenamefont {Retzker}, \citenamefont {McGuinness},\ and\
  \citenamefont {Jelezko}}]{schmitt_submillihertz_2017}%
  \BibitemOpen
  \bibfield  {author} {\bibinfo {author} {\bibfnamefont {S.}~\bibnamefont
  {Schmitt}}, \bibinfo {author} {\bibfnamefont {T.}~\bibnamefont {Gefen}},
  \bibinfo {author} {\bibfnamefont {F.~M.}\ \bibnamefont {Stürner}}, \bibinfo
  {author} {\bibfnamefont {T.}~\bibnamefont {Unden}}, \bibinfo {author}
  {\bibfnamefont {G.}~\bibnamefont {Wolff}}, \bibinfo {author} {\bibfnamefont
  {C.}~\bibnamefont {Müller}}, \bibinfo {author} {\bibfnamefont
  {J.}~\bibnamefont {Scheuer}}, \bibinfo {author} {\bibfnamefont
  {B.}~\bibnamefont {Naydenov}}, \bibinfo {author} {\bibfnamefont
  {M.}~\bibnamefont {Markham}}, \bibinfo {author} {\bibfnamefont
  {S.}~\bibnamefont {Pezzagna}}, \bibinfo {author} {\bibfnamefont
  {J.}~\bibnamefont {Meijer}}, \bibinfo {author} {\bibfnamefont
  {I.}~\bibnamefont {Schwarz}}, \bibinfo {author} {\bibfnamefont {M.~B.}\
  \bibnamefont {Plenio}}, \bibinfo {author} {\bibfnamefont {A.}~\bibnamefont
  {Retzker}}, \bibinfo {author} {\bibfnamefont {L.~P.}\ \bibnamefont
  {McGuinness}},\ and\ \bibinfo {author} {\bibfnamefont {F.}~\bibnamefont
  {Jelezko}},\ }\bibfield  {title} {\bibinfo {title} {Submillihertz magnetic
  spectroscopy performed with a nanoscale quantum sensor},\ }\href
  {https://doi.org/10.1126/science.aam5532} {\bibfield  {journal} {\bibinfo
  {journal} {Science}\ }\textbf {\bibinfo {volume} {356}},\ \bibinfo {pages}
  {832} (\bibinfo {year} {2017})}\BibitemShut {NoStop}%
\bibitem [{\citenamefont {Genov}\ \emph {et~al.}(2019)\citenamefont {Genov},
  \citenamefont {Aharon}, \citenamefont {Jelezko},\ and\ \citenamefont
  {Retzker}}]{genov_mixed_2019}%
  \BibitemOpen
  \bibfield  {author} {\bibinfo {author} {\bibfnamefont {G.~T.}\ \bibnamefont
  {Genov}}, \bibinfo {author} {\bibfnamefont {N.}~\bibnamefont {Aharon}},
  \bibinfo {author} {\bibfnamefont {F.}~\bibnamefont {Jelezko}},\ and\ \bibinfo
  {author} {\bibfnamefont {A.}~\bibnamefont {Retzker}},\ }\bibfield  {title}
  {\bibinfo {title} {Mixed dynamical decoupling},\ }\href
  {https://doi.org/10.1088/2058-9565/ab2afd} {\bibfield  {journal} {\bibinfo
  {journal} {Quantum Science and Technology}\ }\textbf {\bibinfo {volume}
  {4}},\ \bibinfo {pages} {035010} (\bibinfo {year} {2019})}\BibitemShut
  {NoStop}%
\bibitem [{\citenamefont {Ban}(1998)}]{ban_photon-echo_1998}%
  \BibitemOpen
  \bibfield  {author} {\bibinfo {author} {\bibfnamefont {M.}~\bibnamefont
  {Ban}},\ }\bibfield  {title} {\bibinfo {title} {Photon-echo technique for
  reducing the decoherence of a quantum bit},\ }\href
  {https://doi.org/10.1080/09500349808231241} {\bibfield  {journal} {\bibinfo
  {journal} {J. Mod. Opt.}\ }\textbf {\bibinfo {volume} {45}},\ \bibinfo
  {pages} {2315} (\bibinfo {year} {1998})}\BibitemShut {NoStop}%
\bibitem [{\citenamefont {Ozeri}(2013)}]{ozeri_heisenberg_2013}%
  \BibitemOpen
  \bibfield  {author} {\bibinfo {author} {\bibfnamefont {R.}~\bibnamefont
  {Ozeri}},\ }\bibfield  {title} {\bibinfo {title} {Heisenberg limited
  metrology using {Quantum} {Error}-{Correction} {Codes}},\ }\href
  {http://arxiv.org/abs/1310.3432} {\bibfield  {journal} {\bibinfo  {journal}
  {arXiv:1310.3432 [quant-ph]}\ } (\bibinfo {year} {2013})}\BibitemShut
  {NoStop}%
\bibitem [{\citenamefont {Kessler}\ \emph {et~al.}(2014)\citenamefont
  {Kessler}, \citenamefont {Lovchinsky}, \citenamefont {Sushkov},\ and\
  \citenamefont {Lukin}}]{kessler_quantum_2014}%
  \BibitemOpen
  \bibfield  {author} {\bibinfo {author} {\bibfnamefont {E.~M.}\ \bibnamefont
  {Kessler}}, \bibinfo {author} {\bibfnamefont {I.}~\bibnamefont {Lovchinsky}},
  \bibinfo {author} {\bibfnamefont {A.~O.}\ \bibnamefont {Sushkov}},\ and\
  \bibinfo {author} {\bibfnamefont {M.~D.}\ \bibnamefont {Lukin}},\ }\bibfield
  {title} {\bibinfo {title} {Quantum {Error} {Correction} for {Metrology}},\
  }\href {https://doi.org/10.1103/PhysRevLett.112.150802} {\bibfield  {journal}
  {\bibinfo  {journal} {Phys. Rev. Lett.}\ }\textbf {\bibinfo {volume} {112}},\
  \bibinfo {pages} {150802} (\bibinfo {year} {2014})}\BibitemShut {NoStop}%
\bibitem [{\citenamefont {D{\"{u}}r}\ \emph {et~al.}(2014)\citenamefont
  {D{\"{u}}r}, \citenamefont {Skotiniotis}, \citenamefont {Fr{\"{o}}wis},\ and\
  \citenamefont {Kraus}}]{dur_improved_2014}%
  \BibitemOpen
  \bibfield  {author} {\bibinfo {author} {\bibfnamefont {W.}~\bibnamefont
  {D{\"{u}}r}}, \bibinfo {author} {\bibfnamefont {M.}~\bibnamefont
  {Skotiniotis}}, \bibinfo {author} {\bibfnamefont {F.}~\bibnamefont
  {Fr{\"{o}}wis}},\ and\ \bibinfo {author} {\bibfnamefont {B.}~\bibnamefont
  {Kraus}},\ }\bibfield  {title} {\bibinfo {title} {Improved {Quantum}
  {Metrology} {Using} {Quantum} {Error} {Correction}},\ }\href
  {https://doi.org/10.1103/PhysRevLett.112.080801} {\bibfield  {journal}
  {\bibinfo  {journal} {Phys. Rev. Lett.}\ }\textbf {\bibinfo {volume} {112}},\
  \bibinfo {pages} {080801} (\bibinfo {year} {2014})}\BibitemShut {NoStop}%
\bibitem [{\citenamefont {Arrad}\ \emph {et~al.}(2014)\citenamefont {Arrad},
  \citenamefont {Vinkler}, \citenamefont {Aharonov},\ and\ \citenamefont
  {Retzker}}]{arrad_increasing_2014}%
  \BibitemOpen
  \bibfield  {author} {\bibinfo {author} {\bibfnamefont {G.}~\bibnamefont
  {Arrad}}, \bibinfo {author} {\bibfnamefont {Y.}~\bibnamefont {Vinkler}},
  \bibinfo {author} {\bibfnamefont {D.}~\bibnamefont {Aharonov}},\ and\
  \bibinfo {author} {\bibfnamefont {A.}~\bibnamefont {Retzker}},\ }\bibfield
  {title} {\bibinfo {title} {Increasing {Sensing} {Resolution} with {Error}
  {Correction}},\ }\href {https://doi.org/10.1103/PhysRevLett.112.150801}
  {\bibfield  {journal} {\bibinfo  {journal} {Phys. Rev. Lett.}\ }\textbf
  {\bibinfo {volume} {112}},\ \bibinfo {pages} {150801} (\bibinfo {year}
  {2014})}\BibitemShut {NoStop}%
\bibitem [{\citenamefont {Herrera-Mart{\'\i}}\ \emph
  {et~al.}(2015)\citenamefont {Herrera-Mart{\'\i}}, \citenamefont {Gefen},
  \citenamefont {Aharonov}, \citenamefont {Katz},\ and\ \citenamefont
  {Retzker}}]{herrera-marti_quantum_2015}%
  \BibitemOpen
  \bibfield  {author} {\bibinfo {author} {\bibfnamefont {D.~A.}\ \bibnamefont
  {Herrera-Mart{\'\i}}}, \bibinfo {author} {\bibfnamefont {T.}~\bibnamefont
  {Gefen}}, \bibinfo {author} {\bibfnamefont {D.}~\bibnamefont {Aharonov}},
  \bibinfo {author} {\bibfnamefont {N.}~\bibnamefont {Katz}},\ and\ \bibinfo
  {author} {\bibfnamefont {A.}~\bibnamefont {Retzker}},\ }\bibfield  {title}
  {\bibinfo {title} {Quantum {Error}-{Correction}-{Enhanced} {Magnetometer}
  {Overcoming} the {Limit} {Imposed} by {Relaxation}},\ }\href
  {https://doi.org/10.1103/PhysRevLett.115.200501} {\bibfield  {journal}
  {\bibinfo  {journal} {Phys. Rev. Lett.}\ }\textbf {\bibinfo {volume} {115}},\
  \bibinfo {pages} {200501} (\bibinfo {year} {2015})}\BibitemShut {NoStop}%
\bibitem [{\citenamefont {Plenio}\ and\ \citenamefont
  {Huelga}(2016)}]{plenio_sensing_2016}%
  \BibitemOpen
  \bibfield  {author} {\bibinfo {author} {\bibfnamefont {M.~B.}\ \bibnamefont
  {Plenio}}\ and\ \bibinfo {author} {\bibfnamefont {S.~F.}\ \bibnamefont
  {Huelga}},\ }\bibfield  {title} {\bibinfo {title} {Sensing in the presence of
  an observed environment},\ }\href
  {https://doi.org/10.1103/PhysRevA.93.032123} {\bibfield  {journal} {\bibinfo
  {journal} {Phys. Rev. A}\ }\textbf {\bibinfo {volume} {93}},\ \bibinfo
  {pages} {032123} (\bibinfo {year} {2016})}\BibitemShut {NoStop}%
\bibitem [{\citenamefont {Gefen}\ \emph {et~al.}(2016)\citenamefont {Gefen},
  \citenamefont {Herrera-Mart{\'\i}},\ and\ \citenamefont
  {Retzker}}]{gefen_parameter_2016}%
  \BibitemOpen
  \bibfield  {author} {\bibinfo {author} {\bibfnamefont {T.}~\bibnamefont
  {Gefen}}, \bibinfo {author} {\bibfnamefont {D.~A.}\ \bibnamefont
  {Herrera-Mart{\'\i}}},\ and\ \bibinfo {author} {\bibfnamefont
  {A.}~\bibnamefont {Retzker}},\ }\bibfield  {title} {\bibinfo {title}
  {Parameter estimation with efficient photodetectors},\ }\href
  {https://doi.org/10.1103/PhysRevA.93.032133} {\bibfield  {journal} {\bibinfo
  {journal} {Phys. Rev. A}\ }\textbf {\bibinfo {volume} {93}},\ \bibinfo
  {pages} {032133} (\bibinfo {year} {2016})}\BibitemShut {NoStop}%
\bibitem [{\citenamefont {Reiter}\ \emph {et~al.}(2017)\citenamefont {Reiter},
  \citenamefont {Sørensen}, \citenamefont {Zoller},\ and\ \citenamefont
  {Muschik}}]{reiter_dissipative_2017}%
  \BibitemOpen
  \bibfield  {author} {\bibinfo {author} {\bibfnamefont {F.}~\bibnamefont
  {Reiter}}, \bibinfo {author} {\bibfnamefont {A.~S.}\ \bibnamefont
  {Sørensen}}, \bibinfo {author} {\bibfnamefont {P.}~\bibnamefont {Zoller}},\
  and\ \bibinfo {author} {\bibfnamefont {C.~A.}\ \bibnamefont {Muschik}},\
  }\bibfield  {title} {\bibinfo {title} {Dissipative quantum error correction
  and application to quantum sensing with trapped ions},\ }\href
  {https://doi.org/10.1038/s41467-017-01895-5} {\bibfield  {journal} {\bibinfo
  {journal} {Nat. Commun.}\ }\textbf {\bibinfo {volume} {8}},\ \bibinfo {pages}
  {1822} (\bibinfo {year} {2017})}\BibitemShut {NoStop}%
\bibitem [{\citenamefont {Matsuzaki}\ and\ \citenamefont
  {Benjamin}(2017)}]{matsuzaki_magnetic-field_2017}%
  \BibitemOpen
  \bibfield  {author} {\bibinfo {author} {\bibfnamefont {Y.}~\bibnamefont
  {Matsuzaki}}\ and\ \bibinfo {author} {\bibfnamefont {S.}~\bibnamefont
  {Benjamin}},\ }\bibfield  {title} {\bibinfo {title} {Magnetic-field sensing
  with quantum error detection under the effect of energy relaxation},\ }\href
  {https://doi.org/10.1103/PhysRevA.95.032303} {\bibfield  {journal} {\bibinfo
  {journal} {Phys. Rev. A}\ }\textbf {\bibinfo {volume} {95}},\ \bibinfo
  {pages} {032303} (\bibinfo {year} {2017})}\BibitemShut {NoStop}%
\bibitem [{\citenamefont {Unden}\ \emph {et~al.}(2016)\citenamefont {Unden},
  \citenamefont {Balasubramanian}, \citenamefont {Louzon}, \citenamefont
  {Vinkler}, \citenamefont {Plenio}, \citenamefont {Markham}, \citenamefont
  {Twitchen}, \citenamefont {Stacey}, \citenamefont {Lovchinsky}, \citenamefont
  {Sushkov}, \citenamefont {Lukin}, \citenamefont {Retzker}, \citenamefont
  {Naydenov}, \citenamefont {McGuinness},\ and\ \citenamefont
  {Jelezko}}]{unden_quantum_2016}%
  \BibitemOpen
  \bibfield  {author} {\bibinfo {author} {\bibfnamefont {T.}~\bibnamefont
  {Unden}}, \bibinfo {author} {\bibfnamefont {P.}~\bibnamefont
  {Balasubramanian}}, \bibinfo {author} {\bibfnamefont {D.}~\bibnamefont
  {Louzon}}, \bibinfo {author} {\bibfnamefont {Y.}~\bibnamefont {Vinkler}},
  \bibinfo {author} {\bibfnamefont {M.~B.}\ \bibnamefont {Plenio}}, \bibinfo
  {author} {\bibfnamefont {M.}~\bibnamefont {Markham}}, \bibinfo {author}
  {\bibfnamefont {D.}~\bibnamefont {Twitchen}}, \bibinfo {author}
  {\bibfnamefont {A.}~\bibnamefont {Stacey}}, \bibinfo {author} {\bibfnamefont
  {I.}~\bibnamefont {Lovchinsky}}, \bibinfo {author} {\bibfnamefont {A.~O.}\
  \bibnamefont {Sushkov}}, \bibinfo {author} {\bibfnamefont {M.~D.}\
  \bibnamefont {Lukin}}, \bibinfo {author} {\bibfnamefont {A.}~\bibnamefont
  {Retzker}}, \bibinfo {author} {\bibfnamefont {B.}~\bibnamefont {Naydenov}},
  \bibinfo {author} {\bibfnamefont {L.~P.}\ \bibnamefont {McGuinness}},\ and\
  \bibinfo {author} {\bibfnamefont {F.}~\bibnamefont {Jelezko}},\ }\bibfield
  {title} {\bibinfo {title} {Quantum {Metrology} {Enhanced} by {Repetitive}
  {Quantum} {Error} {Correction}},\ }\href
  {https://doi.org/10.1103/PhysRevLett.116.230502} {\bibfield  {journal}
  {\bibinfo  {journal} {Phys. Rev. Lett.}\ }\textbf {\bibinfo {volume} {116}},\
  \bibinfo {pages} {230502} (\bibinfo {year} {2016})}\BibitemShut {NoStop}%
\bibitem [{\citenamefont {Cohen}\ \emph {et~al.}(2016)\citenamefont {Cohen},
  \citenamefont {Pilnyak}, \citenamefont {Istrati}, \citenamefont {Retzker},\
  and\ \citenamefont {Eisenberg}}]{cohen_demonstration_2016}%
  \BibitemOpen
  \bibfield  {author} {\bibinfo {author} {\bibfnamefont {L.}~\bibnamefont
  {Cohen}}, \bibinfo {author} {\bibfnamefont {Y.}~\bibnamefont {Pilnyak}},
  \bibinfo {author} {\bibfnamefont {D.}~\bibnamefont {Istrati}}, \bibinfo
  {author} {\bibfnamefont {A.}~\bibnamefont {Retzker}},\ and\ \bibinfo {author}
  {\bibfnamefont {H.~S.}\ \bibnamefont {Eisenberg}},\ }\bibfield  {title}
  {\bibinfo {title} {Demonstration of a quantum error correction for enhanced
  sensitivity of photonic measurements},\ }\href
  {https://doi.org/10.1103/PhysRevA.94.012324} {\bibfield  {journal} {\bibinfo
  {journal} {Phys. Rev. A}\ }\textbf {\bibinfo {volume} {94}},\ \bibinfo
  {pages} {012324} (\bibinfo {year} {2016})}\BibitemShut {NoStop}%
\bibitem [{\citenamefont {Bergmann}\ and\ \citenamefont {van
  Loock}(2016)}]{bergmann:2016}%
  \BibitemOpen
  \bibfield  {author} {\bibinfo {author} {\bibfnamefont {M.}~\bibnamefont
  {Bergmann}}\ and\ \bibinfo {author} {\bibfnamefont {P.}~\bibnamefont {van
  Loock}},\ }\bibfield  {title} {\bibinfo {title} {Quantum error correction
  against photon loss using noon states},\ }\href
  {https://doi.org/10.1103/PhysRevA.94.012311} {\bibfield  {journal} {\bibinfo
  {journal} {Phys. Rev. A}\ }\textbf {\bibinfo {volume} {94}},\ \bibinfo
  {pages} {012311} (\bibinfo {year} {2016})}\BibitemShut {NoStop}%
\bibitem [{\citenamefont {Sekatski}\ \emph {et~al.}(2017)\citenamefont
  {Sekatski}, \citenamefont {Skotiniotis}, \citenamefont
  {Ko{\l{}}ody{\'{n}}ski},\ and\ \citenamefont {D{\"{u}}r}}]{sekatski:2017}%
  \BibitemOpen
  \bibfield  {author} {\bibinfo {author} {\bibfnamefont {P.}~\bibnamefont
  {Sekatski}}, \bibinfo {author} {\bibfnamefont {M.}~\bibnamefont
  {Skotiniotis}}, \bibinfo {author} {\bibfnamefont {J.}~\bibnamefont
  {Ko{\l{}}ody{\'{n}}ski}},\ and\ \bibinfo {author} {\bibfnamefont
  {W.}~\bibnamefont {D{\"{u}}r}},\ }\bibfield  {title} {\bibinfo {title}
  {Quantum metrology with full and fast quantum control},\ }\href
  {https://doi.org/10.22331/q-2017-09-06-27} {\bibfield  {journal} {\bibinfo
  {journal} {{Quantum}}\ }\textbf {\bibinfo {volume} {1}},\ \bibinfo {pages}
  {27} (\bibinfo {year} {2017})}\BibitemShut {NoStop}%
\bibitem [{\citenamefont {Zhou}\ \emph {et~al.}(2018)\citenamefont {Zhou},
  \citenamefont {Zhang}, \citenamefont {Preskill},\ and\ \citenamefont
  {Jiang}}]{zhou:2018}%
  \BibitemOpen
  \bibfield  {author} {\bibinfo {author} {\bibfnamefont {S.}~\bibnamefont
  {Zhou}}, \bibinfo {author} {\bibfnamefont {M.}~\bibnamefont {Zhang}},
  \bibinfo {author} {\bibfnamefont {J.}~\bibnamefont {Preskill}},\ and\
  \bibinfo {author} {\bibfnamefont {L.}~\bibnamefont {Jiang}},\ }\bibfield
  {title} {\bibinfo {title} {Achieving the heisenberg limit in quantum
  metrology using quantum error correction},\ }\href@noop {} {\bibfield
  {journal} {\bibinfo  {journal} {Nat. Commun.}\ }\textbf {\bibinfo {volume}
  {9}},\ \bibinfo {pages} {78} (\bibinfo {year} {2018})}\BibitemShut {NoStop}%
\bibitem [{\citenamefont {Demkowicz-Dobrza{\'{n}}ski}\ \emph
  {et~al.}(2017)\citenamefont {Demkowicz-Dobrza{\'{n}}ski}, \citenamefont
  {Czajkowski},\ and\ \citenamefont {Sekatski}}]{demkowicz:2017}%
  \BibitemOpen
  \bibfield  {author} {\bibinfo {author} {\bibfnamefont {R.}~\bibnamefont
  {Demkowicz-Dobrza{\'{n}}ski}}, \bibinfo {author} {\bibfnamefont
  {J.}~\bibnamefont {Czajkowski}},\ and\ \bibinfo {author} {\bibfnamefont
  {P.}~\bibnamefont {Sekatski}},\ }\bibfield  {title} {\bibinfo {title}
  {Adaptive quantum metrology under general markovian noise},\ }\href
  {https://doi.org/10.1103/PhysRevX.7.041009} {\bibfield  {journal} {\bibinfo
  {journal} {Phys. Rev. X}\ }\textbf {\bibinfo {volume} {7}},\ \bibinfo {pages}
  {041009} (\bibinfo {year} {2017})}\BibitemShut {NoStop}%
\bibitem [{\citenamefont {Layden}\ and\ \citenamefont
  {Cappellaro}(2018)}]{layden:2018}%
  \BibitemOpen
  \bibfield  {author} {\bibinfo {author} {\bibfnamefont {D.}~\bibnamefont
  {Layden}}\ and\ \bibinfo {author} {\bibfnamefont {P.}~\bibnamefont
  {Cappellaro}},\ }\bibfield  {title} {\bibinfo {title} {Spatial noise
  filtering through error correction for quantum sensing},\ }\href@noop {}
  {\bibfield  {journal} {\bibinfo  {journal} {npj Quantum Inf.}\ }\textbf
  {\bibinfo {volume} {4}},\ \bibinfo {pages} {30} (\bibinfo {year}
  {2018})}\BibitemShut {NoStop}%
\bibitem [{\citenamefont {Layden}\ \emph {et~al.}(2019)\citenamefont {Layden},
  \citenamefont {Zhou}, \citenamefont {Cappellaro},\ and\ \citenamefont
  {Jiang}}]{layden:2019}%
  \BibitemOpen
  \bibfield  {author} {\bibinfo {author} {\bibfnamefont {D.}~\bibnamefont
  {Layden}}, \bibinfo {author} {\bibfnamefont {S.}~\bibnamefont {Zhou}},
  \bibinfo {author} {\bibfnamefont {P.}~\bibnamefont {Cappellaro}},\ and\
  \bibinfo {author} {\bibfnamefont {L.}~\bibnamefont {Jiang}},\ }\bibfield
  {title} {\bibinfo {title} {Ancilla-free quantum error correction codes for
  quantum metrology},\ }\href {https://doi.org/10.1103/PhysRevLett.122.040502}
  {\bibfield  {journal} {\bibinfo  {journal} {Phys. Rev. Lett.}\ }\textbf
  {\bibinfo {volume} {122}},\ \bibinfo {pages} {040502} (\bibinfo {year}
  {2019})}\BibitemShut {NoStop}%
\bibitem [{\citenamefont {G{\'{o}}recki}\ \emph {et~al.}(2020)\citenamefont
  {G{\'{o}}recki}, \citenamefont {Zhou}, \citenamefont {Jiang},\ and\
  \citenamefont {Demkowicz-Dobrza{\'{n}}ski}}]{gorecki:2020}%
  \BibitemOpen
  \bibfield  {author} {\bibinfo {author} {\bibfnamefont {W.}~\bibnamefont
  {G{\'{o}}recki}}, \bibinfo {author} {\bibfnamefont {S.}~\bibnamefont {Zhou}},
  \bibinfo {author} {\bibfnamefont {L.}~\bibnamefont {Jiang}},\ and\ \bibinfo
  {author} {\bibfnamefont {R.}~\bibnamefont {Demkowicz-Dobrza{\'{n}}ski}},\
  }\bibfield  {title} {\bibinfo {title} {Optimal probes and error-correction
  schemes in multi-parameter quantum metrology},\ }\href
  {https://doi.org/10.22331/q-2020-07-02-288} {\bibfield  {journal} {\bibinfo
  {journal} {{Quantum}}\ }\textbf {\bibinfo {volume} {4}},\ \bibinfo {pages}
  {288} (\bibinfo {year} {2020})}\BibitemShut {NoStop}%
\bibitem [{\citenamefont {Lidar}\ and\ \citenamefont
  {Brun}(2013)}]{lidar:2013}%
  \BibitemOpen
  \bibfield  {author} {\bibinfo {author} {\bibfnamefont {D.~A.}\ \bibnamefont
  {Lidar}}\ and\ \bibinfo {author} {\bibfnamefont {T.~A.}\ \bibnamefont
  {Brun}},\ }\href@noop {} {\emph {\bibinfo {title} {Quantum Error
  Correction}}}\ (\bibinfo  {publisher} {Cambridge University Press},\ \bibinfo
  {year} {2013})\BibitemShut {NoStop}%
\bibitem [{Note1()}]{Note1}%
  \BibitemOpen
  \bibinfo {note} {Instead of considering $H\propto \sigma _z$ and $L_\protect
  \mathrm {err}\propto \sigma _x$, some works looked at the problem from the
  opposite point of view by taking $H\propto \sigma _x$ and $L_\protect \mathrm
  {err}\propto \sigma _z$. This only changes the codewords used by the sensing
  protocol but not the conclusions of our work since the signal and the noise
  remain perpendicular in both cases.}\BibitemShut {Stop}%
\bibitem [{\citenamefont {Ramsey}(1950)}]{ramsey_molecular_1950}%
  \BibitemOpen
  \bibfield  {author} {\bibinfo {author} {\bibfnamefont {N.~F.}\ \bibnamefont
  {Ramsey}},\ }\bibfield  {title} {\bibinfo {title} {A {Molecular} {Beam}
  {Resonance} {Method} with {Separated} {Oscillating} {Fields}},\ }\href
  {https://doi.org/10.1103/PhysRev.78.695} {\bibfield  {journal} {\bibinfo
  {journal} {Phys. Rev.}\ }\textbf {\bibinfo {volume} {78}},\ \bibinfo {pages}
  {695} (\bibinfo {year} {1950})}\BibitemShut {NoStop}%
\bibitem [{Note2()}]{Note2}%
  \BibitemOpen
  \bibinfo {note} {See Supplemental Material.}\BibitemShut {Stop}%
\bibitem [{\citenamefont {Wallman}\ and\ \citenamefont
  {Emerson}(2016)}]{wallman_noise_2016}%
  \BibitemOpen
  \bibfield  {author} {\bibinfo {author} {\bibfnamefont {J.~J.}\ \bibnamefont
  {Wallman}}\ and\ \bibinfo {author} {\bibfnamefont {J.}~\bibnamefont
  {Emerson}},\ }\bibfield  {title} {\bibinfo {title} {Noise tailoring for
  scalable quantum computation via randomized compiling},\ }\href
  {https://doi.org/10.1103/PhysRevA.94.052325} {\bibfield  {journal} {\bibinfo
  {journal} {Phys. Rev. A}\ }\textbf {\bibinfo {volume} {94}},\ \bibinfo
  {pages} {052325} (\bibinfo {year} {2016})}\BibitemShut {NoStop}%
\bibitem [{\citenamefont {Harper}\ \emph {et~al.}(2020)\citenamefont {Harper},
  \citenamefont {Flammia},\ and\ \citenamefont
  {Wallman}}]{harper_efficient_2020}%
  \BibitemOpen
  \bibfield  {author} {\bibinfo {author} {\bibfnamefont {R.}~\bibnamefont
  {Harper}}, \bibinfo {author} {\bibfnamefont {S.~T.}\ \bibnamefont
  {Flammia}},\ and\ \bibinfo {author} {\bibfnamefont {J.~J.}\ \bibnamefont
  {Wallman}},\ }\bibfield  {title} {\bibinfo {title} {Efficient learning of
  quantum noise},\ }\bibfield  {journal} {\bibinfo  {journal} {Nat. Phys.}\
  }\textbf {\bibinfo {volume} {16}},\ \href
  {https://doi.org/10.1038/s41567-020-0992-8} {10.1038/s41567-020-0992-8}
  (\bibinfo {year} {2020})\BibitemShut {NoStop}%
\bibitem [{\citenamefont {Ware}\ \emph {et~al.}(2021)\citenamefont {Ware},
  \citenamefont {Ribeill}, \citenamefont {Rist{\`e}}, \citenamefont {Ryan},
  \citenamefont {Johnson},\ and\ \citenamefont
  {da~Silva}}]{ware_experimental_2021}%
  \BibitemOpen
  \bibfield  {author} {\bibinfo {author} {\bibfnamefont {M.}~\bibnamefont
  {Ware}}, \bibinfo {author} {\bibfnamefont {G.}~\bibnamefont {Ribeill}},
  \bibinfo {author} {\bibfnamefont {D.}~\bibnamefont {Rist{\`e}}}, \bibinfo
  {author} {\bibfnamefont {C.~A.}\ \bibnamefont {Ryan}}, \bibinfo {author}
  {\bibfnamefont {B.}~\bibnamefont {Johnson}},\ and\ \bibinfo {author}
  {\bibfnamefont {M.~P.}\ \bibnamefont {da~Silva}},\ }\bibfield  {title}
  {\bibinfo {title} {Experimental {Pauli}-frame randomization on a
  superconducting qubit},\ }\href {https://doi.org/10.1103/PhysRevA.103.042604}
  {\bibfield  {journal} {\bibinfo  {journal} {Phys. Rev. A}\ }\textbf {\bibinfo
  {volume} {103}},\ \bibinfo {pages} {042604} (\bibinfo {year}
  {2021})}\BibitemShut {NoStop}%
\bibitem [{\citenamefont {Cramér}(1999)}]{cramer_mathematical_1999}%
  \BibitemOpen
  \bibfield  {author} {\bibinfo {author} {\bibfnamefont {H.}~\bibnamefont
  {Cramér}},\ }\href {https://doi.org/10.2307/j.ctt1bpm9r4} {\emph {\bibinfo
  {title} {Mathematical {Methods} of {Statistics} ({PMS}-9)}}}\ (\bibinfo
  {publisher} {Princeton University Press},\ \bibinfo {year}
  {1999})\BibitemShut {NoStop}%
\bibitem [{\citenamefont {Huelga}\ \emph {et~al.}(1997)\citenamefont {Huelga},
  \citenamefont {Macchiavello}, \citenamefont {Pellizzari}, \citenamefont
  {Ekert}, \citenamefont {Plenio},\ and\ \citenamefont
  {Cirac}}]{huelga_improvement_1997}%
  \BibitemOpen
  \bibfield  {author} {\bibinfo {author} {\bibfnamefont {S.~F.}\ \bibnamefont
  {Huelga}}, \bibinfo {author} {\bibfnamefont {C.}~\bibnamefont
  {Macchiavello}}, \bibinfo {author} {\bibfnamefont {T.}~\bibnamefont
  {Pellizzari}}, \bibinfo {author} {\bibfnamefont {A.~K.}\ \bibnamefont
  {Ekert}}, \bibinfo {author} {\bibfnamefont {M.~B.}\ \bibnamefont {Plenio}},\
  and\ \bibinfo {author} {\bibfnamefont {J.~I.}\ \bibnamefont {Cirac}},\
  }\bibfield  {title} {\bibinfo {title} {Improvement of {Frequency} {Standards}
  with {Quantum} {Entanglement}},\ }\href
  {https://doi.org/10.1103/PhysRevLett.79.3865} {\bibfield  {journal} {\bibinfo
   {journal} {Phys. Rev. Lett.}\ }\textbf {\bibinfo {volume} {79}},\ \bibinfo
  {pages} {3865} (\bibinfo {year} {1997})}\BibitemShut {NoStop}%
\bibitem [{\citenamefont {Berry}(1941)}]{berry_accuracy_1941}%
  \BibitemOpen
  \bibfield  {author} {\bibinfo {author} {\bibfnamefont {A.~C.}\ \bibnamefont
  {Berry}},\ }\bibfield  {title} {\bibinfo {title} {The {Accuracy} of the
  {Gaussian} {Approximation} to the {Sum} of {Independent} {Variates}},\ }\href
  {https://doi.org/10.2307/1990053} {\bibfield  {journal} {\bibinfo  {journal}
  {Transactions of the American Mathematical Society}\ }\textbf {\bibinfo
  {volume} {49}},\ \bibinfo {pages} {122} (\bibinfo {year} {1941})},\ \bibinfo
  {note} {publisher: American Mathematical Society}\BibitemShut {NoStop}%
\bibitem [{\citenamefont {Esseen}(1942)}]{esseen_liapunov_1942}%
  \BibitemOpen
  \bibfield  {author} {\bibinfo {author} {\bibfnamefont {C.~G.}\ \bibnamefont
  {Esseen}},\ }\bibfield  {title} {\bibinfo {title} {On the {Liapunov} limit
  error in the theory of probability},\ }\href
  {https://ci.nii.ac.jp/naid/10017470730/} {\bibfield  {journal} {\bibinfo
  {journal} {Ark. Mat. Astr. Fys.}\ }\textbf {\bibinfo {volume} {28}},\
  \bibinfo {pages} {1} (\bibinfo {year} {1942})}\BibitemShut {NoStop}%
\bibitem [{\citenamefont {Flühmann}(2018)}]{fluhmann_sequential_2018}%
  \BibitemOpen
  \bibfield  {author} {\bibinfo {author} {\bibfnamefont {C.}~\bibnamefont
  {Flühmann}},\ }\bibfield  {title} {\bibinfo {title} {Sequential {Modular}
  {Position} and {Momentum} {Measurements} of a {Trapped} {Ion} {Mechanical}
  {Oscillator}},\ }\bibfield  {journal} {\bibinfo  {journal} {Phys. Rev. X}\
  }\textbf {\bibinfo {volume} {8}},\ \href
  {https://doi.org/10.1103/PhysRevX.8.021001} {10.1103/PhysRevX.8.021001}
  (\bibinfo {year} {2018})\BibitemShut {NoStop}%
\bibitem [{\citenamefont {Flühmann}\ \emph {et~al.}(2019)\citenamefont
  {Flühmann}, \citenamefont {Nguyen}, \citenamefont {Marinelli}, \citenamefont
  {Negnevitsky}, \citenamefont {Mehta},\ and\ \citenamefont
  {Home}}]{fluhmann_encoding_2019}%
  \BibitemOpen
  \bibfield  {author} {\bibinfo {author} {\bibfnamefont {C.}~\bibnamefont
  {Flühmann}}, \bibinfo {author} {\bibfnamefont {T.~L.}\ \bibnamefont
  {Nguyen}}, \bibinfo {author} {\bibfnamefont {M.}~\bibnamefont {Marinelli}},
  \bibinfo {author} {\bibfnamefont {V.}~\bibnamefont {Negnevitsky}}, \bibinfo
  {author} {\bibfnamefont {K.}~\bibnamefont {Mehta}},\ and\ \bibinfo {author}
  {\bibfnamefont {J.~P.}\ \bibnamefont {Home}},\ }\bibfield  {title} {\bibinfo
  {title} {Encoding a qubit in a trapped-ion mechanical oscillator},\ }\href
  {https://doi.org/10.1038/s41586-019-0960-6} {\bibfield  {journal} {\bibinfo
  {journal} {Nature}\ }\textbf {\bibinfo {volume} {566}},\ \bibinfo {pages}
  {513} (\bibinfo {year} {2019})}\BibitemShut {NoStop}%
\bibitem [{\citenamefont {Campagne-Ibarcq}\ \emph {et~al.}(2020)\citenamefont
  {Campagne-Ibarcq}, \citenamefont {Eickbusch}, \citenamefont {Touzard},
  \citenamefont {Zalys-Geller}, \citenamefont {Frattini}, \citenamefont
  {Sivak}, \citenamefont {Reinhold}, \citenamefont {Puri}, \citenamefont
  {Shankar}, \citenamefont {Schoelkopf}, \citenamefont {Frunzio}, \citenamefont
  {Mirrahimi},\ and\ \citenamefont {Devoret}}]{campagne-ibarcq_quantum_2020}%
  \BibitemOpen
  \bibfield  {author} {\bibinfo {author} {\bibfnamefont {P.}~\bibnamefont
  {Campagne-Ibarcq}}, \bibinfo {author} {\bibfnamefont {A.}~\bibnamefont
  {Eickbusch}}, \bibinfo {author} {\bibfnamefont {S.}~\bibnamefont {Touzard}},
  \bibinfo {author} {\bibfnamefont {E.}~\bibnamefont {Zalys-Geller}}, \bibinfo
  {author} {\bibfnamefont {N.~E.}\ \bibnamefont {Frattini}}, \bibinfo {author}
  {\bibfnamefont {V.~V.}\ \bibnamefont {Sivak}}, \bibinfo {author}
  {\bibfnamefont {P.}~\bibnamefont {Reinhold}}, \bibinfo {author}
  {\bibfnamefont {S.}~\bibnamefont {Puri}}, \bibinfo {author} {\bibfnamefont
  {S.}~\bibnamefont {Shankar}}, \bibinfo {author} {\bibfnamefont {R.~J.}\
  \bibnamefont {Schoelkopf}}, \bibinfo {author} {\bibfnamefont
  {L.}~\bibnamefont {Frunzio}}, \bibinfo {author} {\bibfnamefont
  {M.}~\bibnamefont {Mirrahimi}},\ and\ \bibinfo {author} {\bibfnamefont
  {M.~H.}\ \bibnamefont {Devoret}},\ }\bibfield  {title} {\bibinfo {title}
  {Quantum error correction of a qubit encoded in grid states of an
  oscillator},\ }\href {https://doi.org/10.1038/s41586-020-2603-3} {\bibfield
  {journal} {\bibinfo  {journal} {Nature}\ }\textbf {\bibinfo {volume} {584}},\
  \bibinfo {pages} {368} (\bibinfo {year} {2020})}\BibitemShut {NoStop}%
\bibitem [{\citenamefont {de~Neeve}\ \emph {et~al.}(2020)\citenamefont
  {de~Neeve}, \citenamefont {Nguyen}, \citenamefont {Behrle},\ and\
  \citenamefont {Home}}]{de_neeve_error_2020}%
  \BibitemOpen
  \bibfield  {author} {\bibinfo {author} {\bibfnamefont {B.}~\bibnamefont
  {de~Neeve}}, \bibinfo {author} {\bibfnamefont {T.~L.}\ \bibnamefont
  {Nguyen}}, \bibinfo {author} {\bibfnamefont {T.}~\bibnamefont {Behrle}},\
  and\ \bibinfo {author} {\bibfnamefont {J.}~\bibnamefont {Home}},\ }\bibfield
  {title} {\bibinfo {title} {Error correction of a logical grid state qubit by
  dissipative pumping},\ }\href {http://arxiv.org/abs/2010.09681} {\bibfield
  {journal} {\bibinfo  {journal} {arXiv:2010.09681 [quant-ph]}\ } (\bibinfo
  {year} {2020})}\BibitemShut {NoStop}%
\bibitem [{\citenamefont {Maiwald}\ \emph {et~al.}(2009)\citenamefont
  {Maiwald}, \citenamefont {Leibfried}, \citenamefont {Britton}, \citenamefont
  {Bergquist}, \citenamefont {Leuchs},\ and\ \citenamefont
  {Wineland}}]{maiwald_stylus_2009}%
  \BibitemOpen
  \bibfield  {author} {\bibinfo {author} {\bibfnamefont {R.}~\bibnamefont
  {Maiwald}}, \bibinfo {author} {\bibfnamefont {D.}~\bibnamefont {Leibfried}},
  \bibinfo {author} {\bibfnamefont {J.}~\bibnamefont {Britton}}, \bibinfo
  {author} {\bibfnamefont {J.~C.}\ \bibnamefont {Bergquist}}, \bibinfo {author}
  {\bibfnamefont {G.}~\bibnamefont {Leuchs}},\ and\ \bibinfo {author}
  {\bibfnamefont {D.~J.}\ \bibnamefont {Wineland}},\ }\bibfield  {title}
  {\bibinfo {title} {Stylus ion trap for enhanced access and sensing},\ }\href
  {https://doi.org/10.1038/nphys1311} {\bibfield  {journal} {\bibinfo
  {journal} {Nat. Phys.}\ }\textbf {\bibinfo {volume} {5}},\ \bibinfo {pages}
  {551} (\bibinfo {year} {2009})}\BibitemShut {NoStop}%
\bibitem [{\citenamefont {Campbell}\ and\ \citenamefont
  {Hamilton}(2017)}]{campbell_rotation_2017}%
  \BibitemOpen
  \bibfield  {author} {\bibinfo {author} {\bibfnamefont {W.~C.}\ \bibnamefont
  {Campbell}}\ and\ \bibinfo {author} {\bibfnamefont {P.}~\bibnamefont
  {Hamilton}},\ }\bibfield  {title} {\bibinfo {title} {Rotation sensing with
  trapped ions},\ }\href {https://doi.org/10.1088/1361-6455/aa5a8f} {\bibfield
  {journal} {\bibinfo  {journal} {J. Phys. B: At., Mol. Opt. Phys.}\ }\textbf
  {\bibinfo {volume} {50}},\ \bibinfo {pages} {064002} (\bibinfo {year}
  {2017})}\BibitemShut {NoStop}%
\bibitem [{\citenamefont {Biercuk}\ \emph {et~al.}(2010)\citenamefont
  {Biercuk}, \citenamefont {Uys}, \citenamefont {Britton}, \citenamefont
  {VanDevender},\ and\ \citenamefont
  {Bollinger}}]{biercuk_ultrasensitive_2010}%
  \BibitemOpen
  \bibfield  {author} {\bibinfo {author} {\bibfnamefont {M.~J.}\ \bibnamefont
  {Biercuk}}, \bibinfo {author} {\bibfnamefont {H.}~\bibnamefont {Uys}},
  \bibinfo {author} {\bibfnamefont {J.~W.}\ \bibnamefont {Britton}}, \bibinfo
  {author} {\bibfnamefont {A.~P.}\ \bibnamefont {VanDevender}},\ and\ \bibinfo
  {author} {\bibfnamefont {J.~J.}\ \bibnamefont {Bollinger}},\ }\bibfield
  {title} {\bibinfo {title} {Ultrasensitive detection of force and displacement
  using trapped ions},\ }\href {https://doi.org/10.1038/nnano.2010.165}
  {\bibfield  {journal} {\bibinfo  {journal} {Nat. Nanotechnol.}\ }\textbf
  {\bibinfo {volume} {5}},\ \bibinfo {pages} {646} (\bibinfo {year}
  {2010})}\BibitemShut {NoStop}%
\bibitem [{\citenamefont {Shaniv}\ and\ \citenamefont
  {Ozeri}(2017)}]{shaniv_quantum_2017}%
  \BibitemOpen
  \bibfield  {author} {\bibinfo {author} {\bibfnamefont {R.}~\bibnamefont
  {Shaniv}}\ and\ \bibinfo {author} {\bibfnamefont {R.}~\bibnamefont {Ozeri}},\
  }\bibfield  {title} {\bibinfo {title} {Quantum lock-in force sensing using
  optical clock {Doppler} velocimetry},\ }\href
  {https://doi.org/10.1038/ncomms14157} {\bibfield  {journal} {\bibinfo
  {journal} {Nat. Commun.}\ }\textbf {\bibinfo {volume} {8}},\ \bibinfo {pages}
  {14157} (\bibinfo {year} {2017})}\BibitemShut {NoStop}%
\bibitem [{\citenamefont {Duivenvoorden}\ \emph {et~al.}()\citenamefont
  {Duivenvoorden}, \citenamefont {Terhal},\ and\ \citenamefont
  {Weigand}}]{duivenvoorden_single-mode_2017}%
  \BibitemOpen
  \bibfield  {author} {\bibinfo {author} {\bibfnamefont {K.}~\bibnamefont
  {Duivenvoorden}}, \bibinfo {author} {\bibfnamefont {B.~M.}\ \bibnamefont
  {Terhal}},\ and\ \bibinfo {author} {\bibfnamefont {D.}~\bibnamefont
  {Weigand}},\ }\bibfield  {title} {\bibinfo {title} {Single-mode displacement
  sensor},\ }\href {https://doi.org/10.1103/PhysRevA.95.012305} {\bibfield
  {journal} {\bibinfo  {journal} {Phys. Rev. A}\ }\textbf {\bibinfo {volume}
  {95}},\ \bibinfo {pages} {012305}}\BibitemShut {NoStop}%
\bibitem [{\citenamefont {Wolf}\ \emph {et~al.}(2019)\citenamefont {Wolf},
  \citenamefont {Shi}, \citenamefont {Heip}, \citenamefont {Gessner},
  \citenamefont {Pezzè}, \citenamefont {Smerzi}, \citenamefont {Schulte},
  \citenamefont {Hammerer},\ and\ \citenamefont
  {Schmidt}}]{wolf_motional_2019}%
  \BibitemOpen
  \bibfield  {author} {\bibinfo {author} {\bibfnamefont {F.}~\bibnamefont
  {Wolf}}, \bibinfo {author} {\bibfnamefont {C.}~\bibnamefont {Shi}}, \bibinfo
  {author} {\bibfnamefont {J.~C.}\ \bibnamefont {Heip}}, \bibinfo {author}
  {\bibfnamefont {M.}~\bibnamefont {Gessner}}, \bibinfo {author} {\bibfnamefont
  {L.}~\bibnamefont {Pezzè}}, \bibinfo {author} {\bibfnamefont
  {A.}~\bibnamefont {Smerzi}}, \bibinfo {author} {\bibfnamefont
  {M.}~\bibnamefont {Schulte}}, \bibinfo {author} {\bibfnamefont
  {K.}~\bibnamefont {Hammerer}},\ and\ \bibinfo {author} {\bibfnamefont
  {P.~O.}\ \bibnamefont {Schmidt}},\ }\bibfield  {title} {\bibinfo {title}
  {Motional {Fock} states for quantum-enhanced amplitude and phase measurements
  with trapped ions},\ }\href {https://doi.org/10.1038/s41467-019-10576-4}
  {\bibfield  {journal} {\bibinfo  {journal} {Nat. Commun.}\ }\textbf {\bibinfo
  {volume} {10}},\ \bibinfo {pages} {2929} (\bibinfo {year}
  {2019})}\BibitemShut {NoStop}%
\bibitem [{\citenamefont {Flühmann}(2019)}]{fluhmann_thesis_2019}%
  \BibitemOpen
  \bibfield  {author} {\bibinfo {author} {\bibfnamefont {C.}~\bibnamefont
  {Flühmann}},\ }\emph {\bibinfo {title} {Encoding a qubit in the motion of a
  trapped ion using superpositions of displaced squeezed states}},\ \href
  {https://doi.org/10.3929/ethz-b-000355836} {\bibinfo {type} {Doctoral
  {Thesis}}},\ \bibinfo  {school} {ETH Zurich} (\bibinfo {year}
  {2019})\BibitemShut {NoStop}%
\bibitem [{\citenamefont {Gottesman}\ \emph {et~al.}(2001)\citenamefont
  {Gottesman}, \citenamefont {Kitaev},\ and\ \citenamefont
  {Preskill}}]{gottesman_encoding_2001}%
  \BibitemOpen
  \bibfield  {author} {\bibinfo {author} {\bibfnamefont {D.}~\bibnamefont
  {Gottesman}}, \bibinfo {author} {\bibfnamefont {A.}~\bibnamefont {Kitaev}},\
  and\ \bibinfo {author} {\bibfnamefont {J.}~\bibnamefont {Preskill}},\
  }\bibfield  {title} {\bibinfo {title} {Encoding a qubit in an oscillator},\
  }\href {https://doi.org/10.1103/PhysRevA.64.012310} {\bibfield  {journal}
  {\bibinfo  {journal} {Phys. Rev. A}\ }\textbf {\bibinfo {volume} {64}},\
  \bibinfo {pages} {012310} (\bibinfo {year} {2001})}\BibitemShut {NoStop}%
\bibitem [{\citenamefont {Liu}\ \emph {et~al.}(2019)\citenamefont {Liu},
  \citenamefont {Yuan}, \citenamefont {Lu},\ and\ \citenamefont
  {Wang}}]{liu_quantum_2019}%
  \BibitemOpen
  \bibfield  {author} {\bibinfo {author} {\bibfnamefont {J.}~\bibnamefont
  {Liu}}, \bibinfo {author} {\bibfnamefont {H.}~\bibnamefont {Yuan}}, \bibinfo
  {author} {\bibfnamefont {X.-M.}\ \bibnamefont {Lu}},\ and\ \bibinfo {author}
  {\bibfnamefont {X.}~\bibnamefont {Wang}},\ }\bibfield  {title} {\bibinfo
  {title} {Quantum {Fisher} information matrix and multiparameter estimation},\
  }\href {https://doi.org/10.1088/1751-8121/ab5d4d} {\bibfield  {journal}
  {\bibinfo  {journal} {J. Phys. A: Math. Theor.}\ }\textbf {\bibinfo {volume}
  {53}},\ \bibinfo {pages} {023001} (\bibinfo {year} {2019})}\BibitemShut
  {NoStop}%
\bibitem [{\citenamefont {Geman}\ \emph {et~al.}(1992)\citenamefont {Geman},
  \citenamefont {Bienenstock},\ and\ \citenamefont
  {Doursat}}]{geman_neural_1992}%
  \BibitemOpen
  \bibfield  {author} {\bibinfo {author} {\bibfnamefont {S.}~\bibnamefont
  {Geman}}, \bibinfo {author} {\bibfnamefont {E.}~\bibnamefont {Bienenstock}},\
  and\ \bibinfo {author} {\bibfnamefont {R.}~\bibnamefont {Doursat}},\
  }\bibfield  {title} {\bibinfo {title} {Neural {Networks} and the
  {Bias}/{Variance} {Dilemma}.},\ }\href
  {http://dblp.uni-trier.de/db/journals/neco/neco4.html#GemanBD92} {\bibfield
  {journal} {\bibinfo  {journal} {Neural Comput.}\ }\textbf {\bibinfo {volume}
  {4}},\ \bibinfo {pages} {1} (\bibinfo {year} {1992})}\BibitemShut {NoStop}%
\bibitem [{\citenamefont {Johansson}\ \emph {et~al.}(2013)\citenamefont
  {Johansson}, \citenamefont {Nation},\ and\ \citenamefont
  {Nori}}]{johansson_qutip_2013}%
  \BibitemOpen
  \bibfield  {author} {\bibinfo {author} {\bibfnamefont {J.~R.}\ \bibnamefont
  {Johansson}}, \bibinfo {author} {\bibfnamefont {P.~D.}\ \bibnamefont
  {Nation}},\ and\ \bibinfo {author} {\bibfnamefont {F.}~\bibnamefont {Nori}},\
  }\bibfield  {title} {\bibinfo {title} {{QuTiP} 2: {A} {Python} framework for
  the dynamics of open quantum systems},\ }\href
  {https://doi.org/https://doi.org/10.1016/j.cpc.2012.11.019} {\bibfield
  {journal} {\bibinfo  {journal} {Comput. Phys. Commun.}\ }\textbf {\bibinfo
  {volume} {184}},\ \bibinfo {pages} {1234 } (\bibinfo {year}
  {2013})}\BibitemShut {NoStop}%
\bibitem [{Note3()}]{Note3}%
  \BibitemOpen
  \bibinfo {note} {We do not compare these functions to their simulated
  counterparts as in Section~\ref {sec:disparity}, because their calculation is
  resource demanding.}\BibitemShut {Stop}%
\end{thebibliography}%


\onecolumngrid
\clearpage

\section*{Supplemental Material} 

In this Supplemental Material, we give detailed derivations of the main formulas and provide justifications to assertions from the paper. In particular, we derive the corrected as well as the uncorrected dynamics and discuss their differences and the validity range of their reduced form. Then we derive the main results about the biasedness condition and the generalization of the problem using quantum channels. Using the Cram\'{e}r-Rao bound we show the biasedness of the naive estimator and determine the minimum detectable frequency variation. Finally, the last two sections are dedicated for numerical comparisons between simulations and analytical solutions.

\section{Solution of the master equation} \label{sec:sol_err_corr}
  
  Above all, we determine the components of the density matrix $\rho$ on which the expectation value $\langle \sigma_x^L \rangle$ depends. It is straightforward to see that $\langle \sigma_x^L \rangle = q + q^* = 2\text{Re}(q)$ where $q=\bra{1}_L\rho\ket{0}_L$. This means that the goal is to solve the master equation for this component only. If both, the initialization and the first Hadamard gate, are assumed to be perfect, the initial value of $q$ is equal to $\half$ and only the free evolution stage changes it. The latter is dictated by the master equation~(3) which in the case of this component can be reduced to a system of eight differential equations involving \,$q=\bra{111}\rho\ket{000}$,\, $e_1=\bra{011}\rho\ket{100}$,\, $e_2=\bra{101}\rho\ket{010}$,\, $e_3=\bra{110}\rho\ket{001}$ and their respective complex conjugates. The entire system is presented in \eqref{eq:entire_system} with the dot denoting the derivative with respect to $\tau$. One can notice that the system is invariant under permutation of the error components $\{e_1,\,e_2,\,e_3\}$ and can be then reduced to a system of four rate equations by defining $e=e_1+e_2+e_3$ (cf. Eq.~\eqref{eq:entire_system}).

  \begin{figure*}[ht!]
    \be \label{eq:entire_system}
      \begin{cases}
        \dot{q}     = (-3i\,\omega-3\,\Gerr)\,q +(\Gerr+\Gqec)(e_1+e_2+e_3) \\
        \dot{e}_1   = \Gerr\,q + (-i\omega-3\,\Gerr-\Gqec)\,e_1 + \Gerr\,(e_2^*+e_3^*) \\
        \dot{e}_2   = \Gerr\,q + (-i\omega-3\,\Gerr-\Gqec)\,e_2 + \Gerr\,(e_1^*+e_3^*) \\
        \dot{e}_3   = \Gerr\,q + (-i\omega-3\,\Gerr-\Gqec)\,e_3 + \Gerr\,(e_1^*+e_2^*) \\
        \dot{e}_1^* =  \Gerr\,q^* + (i\omega-3\,\Gerr-\Gqec)\,e_1^* + \Gerr\,(e_2+e_3)\\
        \dot{e}_2^* =  \Gerr\,q^* + (i\omega-3\,\Gerr-\Gqec)\,e_2^* + \Gerr\,(e_1+e_3) \\
        \dot{e}_3^* =  \Gerr\,q^* + (i\omega-3\,\Gerr-\Gqec)\,e_3^* + \Gerr\,(e_1+e_2) \\
        \dot{q}^*   =  (3i\,\omega-3\,\Gerr)\,q^* +(\Gerr+\Gqec)(e_1^*+e_2^*+e_3^*) 
      \end{cases}	\hspace*{-7.5pt}\rightarrow
      \begin{cases}
        \dot{q} = (-3i\,\omega-3\,\Gerr)\,q +(\Gerr+\Gqec)\,e \\
        \dot{e} = 3\,\Gerr\,q + (-i\omega-3\,\Gerr-\Gqec)\,e + 2\,\Gerr\,e^* \\
        \dot{e}^* =  3\,\Gerr\,q^* + (i\omega-3\,\Gerr-\Gqec)\,e + 2\,\Gerr\,e \\
        \dot{q}^*   =  (3i\,\omega-3\,\Gerr)\,q^* +(\Gerr+\Gqec)\,e^* 
      \end{cases}
    \ee\vspace{-10pt}
  \end{figure*}

  This permutation invariant system leads to a complicated set of rate equations which do not have, due to the presence of complex coefficients, a simple solution. A way to bypass this issue is to consider a slightly simpler problem given by the matrix equation \eqref{eq:simple_system} without the orange terms. This approximation can be understood as following: error components $e$ and $e^*$ of the density matrix $\rho$ can only decay through the error channel into an additional virtual leakage subspace whereas in fact this channel just transforms them into each other. Conversely, replacing the $-3\Gerr\,e$ term by $-\Gerr\,e$ would lead to a closed system where double errors couldn't occur at all. This is a more restrictive assumption and thus was not considered.

  \be \label{eq:simple_system}
		\frac{\n{d}}{\n{d}\tau}
		\begin{pmatrix}
		q \\ e \\ e^* \\ q^*	
		\end{pmatrix}
		=
		\begin{pmatrix}
			-3i\omega-3\Gerr & \Gerr+\Gqec & 0 & 0 \\
			3\Gerr & -i\omega-3\Gerr-\Gqec & {\color{orange}2\Gerr} & 0 \\
			0 & {\color{orange}2\Gerr} & i\omega-3\Gerr-\Gqec & 3\Gerr \\
			0 & 0 & \Gerr+\Gqec & 3i\omega-3\Gerr
		\end{pmatrix}
		\begin{pmatrix}
			q \\ e \\ e^* \\ q^*	
		\end{pmatrix}
	\ee

  The problem together with the assumption explained above comes down to find the eigenvalues and eigenvectors of the upper 2x2 matrix from Eq.~\eqref{eq:simple_system}. They are given as following: 
  \be \label{eq:eigenvals}
    \begin{split}
    &\lambda_\pm = -\half\,\Gqec - 3\,\Gerr - 2 i \omega \pm \half\,\sqrt{D}
    \\[5pt]
    &\vec{v}_\pm \equiv
		\begin{pmatrix}[1.5]
			v_\pm^{q} \\
      v_\pm^{e}
		\end{pmatrix} =
    \begin{pmatrix}[1.5]
			\frac{1}{6\,\Gamma_\n{err}}
			\left(\Gamma_\n{qec} - 2i\,\omega \pm \sqrt{D} \right) \\
      1 
    \end{pmatrix} \\[5pt]
    &\textnormal{with}\qquad\quad D=\Gqec^2+12\,\Gqec\,\Gerr+12\,\Gerr^2 -4i\,\Gqec\,\omega-4\,\omega^2 
    \end{split}
  \ee
  The solution consists of a linear combination of products of each eigenvector with the exponential of the corresponding eigenvalue. The initial value of the problem is $(1/2,0,0,1/2)$. This implies:
  \[
		\begin{pmatrix}[1.5]
			q(\tau) \\
      e(\tau)
		\end{pmatrix} 
		= A\,\bexp^{\lambda_+\,\tau}\,\vec{v}_+ \, + \, B\,\bexp^{\lambda_-\,\tau}\,\vec{v}_-
  \]
  where $A$ and $B$ have to be determined knowing that $q(0)=1/2$ and $e(0)=0$; this leads to the following normalization constants $C_\pm$ from Eq.~(6)
  \be \label{eq:constants}
			C_\pm=\frac{\Gqec}{4 \sqrt{D}} - \frac{i \omega}{2 \sqrt{D}} \pm \frac{1}{4}
  \ee

\section{Validity range of the reduced solution} \label{sec:reduced_sol}

  \begin{SCfigure}[][b!]
    \centering
    \def\svgwidth{.5\linewidth}
    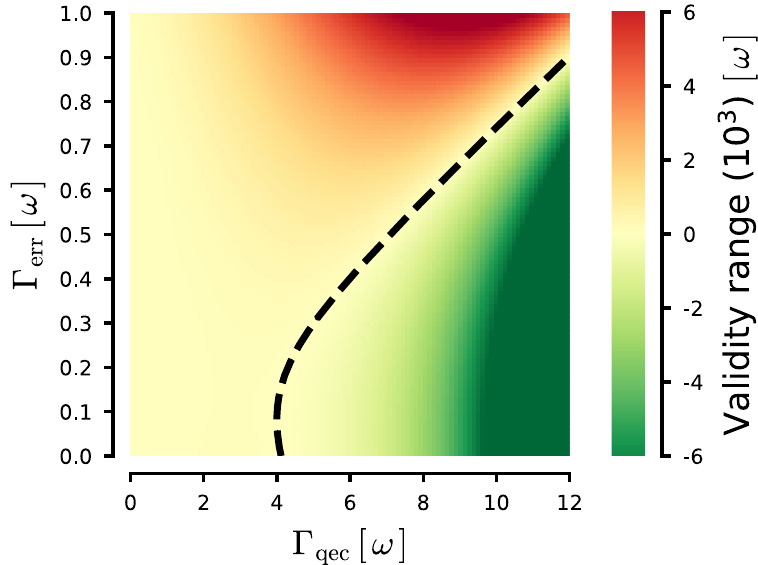
    \caption{\textit{Validity range.} The validity condition (cf. top inequality of Eq.~\eqref{eq:validity_range}) of the approximation given in Eq.~(7) expressed in units of $\omega$. The green region represents the valid parameter space and the red one -- parameters which violate the condition. The dashed line is the limit between these areas.}
    \label{fig:validity_range}
  \end{SCfigure}
  
  In an error-corrected sensing situation, the solution presented in Eq.~(6) does not seem to have an intuitive physical explanation. Indeed, one can hardly distinguish a decaying part caused by errors from a growing part resulting from the error-correction. This is due to the presence of the square-root term in the exponentials. A way to circumvent this is to use an approximation of this term. Let us rewrite it as following:
  \be \label{eq:expansion}
      \sqrt{D} = \Gqec \sqrt{1+ \frac{12\,\Gqec\,\Gerr+12\,\Gerr^2 - 4i\,\Gqec\,\omega-4\,\omega^2}{\Gqec^2}} \,\,.
  \ee
  Here, we have chosen to consider that the correction rate $\Gqec$ dominates over the rest of the terms in the square root. We  expand it up to the second order of the fraction in the square root. The convergence of this expansion is ensured when the absolute value of this ratio is strictly lower than one; stated differently:
  \be \label{eq:validity_range}
  \begin{split}
      &(12\,\Gqec\,\Gerr+12\,\Gerr^2-4\,\omega^2)^2 + 16\,\Gqec^2\,\omega^2 -\Gqec^4 < 0 \\
      &(12\,\Gqec\,\Gerr+12\,\Gerr^2-4\,\omega^2)^2 + 16\,\Gqec^2\,\omega^2 +\Gqec^4 > 0 
  \end{split}
  \ee
  have to be satisfied by the parameters of the system. While the bottom condition is always satisfied if $\Gerr<\omega$, the top one has a nontrivial valid parameter space. Its validity limit is shown in \figref{fig:validity_range} as a dashed line. Parameters inside the red region violate this condition meaning that the expansion does not converge. Assuming that we are in the green region and considering only terms of the order $\cl{O}(1/\Gqec)$, this expansion leads to the expectation value $\langle \sigma_x^L \rangle$ given by Eq.~(7) in the main text. It can then be reduced further to Eq.~(8).

\section{Bias as a function of correction rate} \label{sec:bias_vs_gqec}

\begin{SCfigure}[][h!]
    \centering
    \def\svgwidth{.4\linewidth}
    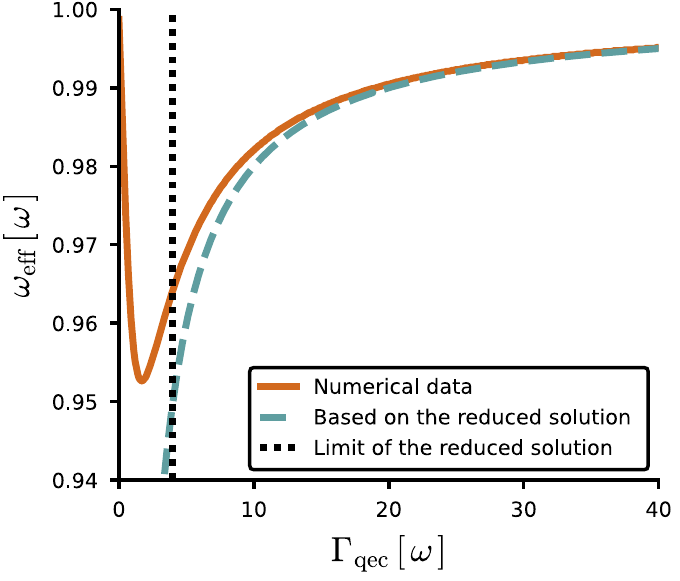
    \caption{ \textit{Bias \texttt{VS} QEC.} Evolution of the effective frequency in terms of correction rate. Numerical data were obtained using a discrete Fourier transform of the proposed solution (cf. Eq.~(6)). The curve is well approximated by the first-order expression of $\oeff$ presented in Eq.~(10). The limit of this approximation is set by inequalities in Eq.~\eqref{eq:validity_range}. The error rate is fixed at $\Gerr=0.1\omega$.}
    \label{fig:bias_vs_qec}
  \end{SCfigure}

  The bias of the frequency is related to all the parameters of the system, namely $\omega$, $\Gerr$ and $\Gqec$. Fixing the two first one and varying the correction rate can help us to determine the value for which the bias becomes negligible. This analysis is shown in \figref{fig:bias_vs_qec}. While the reduced solution given in Eq.~(8) can help to approximate this curve for large $\Gqec$; for low values, the validity conditions~(cf. Eq.~\eqref{eq:validity_range}) are not satisfied anymore. We can observe that after reaching a minimum, the effective frequency starts to increase again. 

  We believe that this arises from the nature of the measurement we are using. Indeed, when the measurement basis is made of logical states, qubits from the erroneous subspace do not corrupt the outcome, such that the only way they bias the measurement is due to multiple bit flips. However, the probability of these events is low, since the qubits population, initially in the logical states, will quickly and equally spread among all the possible states of the Hilbert space. When QEC becomes nontrivial, this probability starts to increase together with the logical population. A non-negligible amount of the latter has, however, spent some time in the erroneous subspace and thus evolved at a different frequency. As we keep increasing $\Gqec$, this time decreases, which consequently reduces the bias.

  \section{Uncorrected sensing} \label{sec:uncorrected_sol}

  In this section we present the analytical result for a faulty but uncorrected Ramsey experiment, in other words we consider $\Gerr > 0$ and $\Gqec=0$. We perform the same analysis as Section~\ref{eq:entire_system}. In this situation, the system of equations reads
  \be \label{eq:entire_system_reduced_err}
      \frac{\n{d}}{\n{d}\tau}
      \begin{pmatrix}
      q \\ e \\ e^* \\ q^*	
      \end{pmatrix}
      =
      \begin{pmatrix}
          -3i\omega-3\Gerr & \Gerr & 0 & 0 \\
          3\Gerr & -i\omega-3\Gerr & 2\Gerr & 0 \\
          0 & 2\Gerr & i\omega-3\Gerr & 3\Gerr \\
          0 & 0 & \Gerr & 3i\omega-3\Gerr
      \end{pmatrix}
      \begin{pmatrix}
          q \\ e \\ e^* \\ q^*	
      \end{pmatrix}_{\,\,.}
  \ee
  This system, unlike in the error-corrected case, can be seamlessly solved using the eigenanalysis of the matrix. The four eigenvalues of the matrix are thus $\lambda_{1,2}=-3\Gerr \pm 3 \sqrt{\Gerr^2-\omega^2}$ \,and  $\lambda_{3,4}=-3\Gerr \pm \sqrt{\Gerr^2-\omega^2}$. Assuming $\Gerr<\omega$, we solve the resulting linear system of equations with the initial value $(1/2,0,0,1/2)$ and obtain the following expression for the expectation value $\langle \sigma_x^L \rangle$ ,
  \be \label{eq:q_3qubits_err}
  \begin{split}
    \langle \sigma_x^L \rangle = 2 \n{Re}(q(\tau))
      = \frac{\omega^2}{D}\,\bexp^{-3\Gerr\tau} \, \cos(3\sqrt{D}\,\tau)
      - \frac{\Gerr^2}{(\sqrt{D})^3}\,\bexp^{-3\Gerr\tau} \, 
      \left( \sqrt{D}\cos^3(\sqrt{D}\,\tau)+\Gerr\sin^3(\sqrt{D}\,\tau)\right) \,\,,
  \end{split}
  \ee
  where $D = \omega^2-\Gerr^2$ is a discriminant. If one supposes further that $\omega \gg \Gerr$, the normalization terms can be expanded
  \be \label{eq:q_3qubits_err_expan}
  \begin{split}
  \langle \sigma_x^L \rangle &=\bexp^{-3\Gerr\tau} \, \cos(3\sqrt{D}\,\tau)\,
                              \left[1 + \left(\frac{\Gerr}{\omega}\right)^2 + 
                              \cl{O}\left(\frac{\Gerr}{\omega}\right)^4 \,
                              \right]- \\
                  &-\bexp^{-3\Gerr\tau} \, \cos^3(\sqrt{D}\,\tau) \,
                              \left[\left(\frac{\Gerr}{\omega}\right)^2 + 
                              \cl{O}\left(\frac{\Gerr}{\omega}\right)^4 \,
                              \right] 
                  -\bexp^{-3\Gerr\tau} \, \sin^3(\sqrt{D}\,\tau) \,
                              \cl{O}\left(\frac{\Gerr}{\omega}\right)^4 \,\, .
  \end{split}
  \ee
  Together with the expansion of the frequency term $\sqrt{D}$, we conclude that, up to second order in $\frac{\Gerr}{\omega}$, the evolution of the expectation value of the logical Pauli-x operator is given by
  \be \label{eq:uncorrected_sol}
    \langle \sigma_x^L \rangle \approx \bexp^{-3\Gerr\tau} \, \cos[\,3\,\omega_\n{err}\,\tau\,] 
    \qquad\n{with}\qquad
    \omega_\n{err} = \omega\,\left(1-\half\frac{\Gerr^2}{\omega^2}+\cl{O}(\frac{\Gerr^4}{\omega^4})\right) \,\,.
  \ee

  We observe that, up to the first order in $\frac{\Gerr}{\omega}$, Eq.~(5) effectively approximates the dynamics of the expectation value $\langle \sigma_x^L \rangle$ in the presence of low bit-flip rate and no correction. Higher-order terms of the expansion introduce a scaling factor $1-\half\frac{\Gerr^2}{\omega^2}$ in the frequency of the measured signal. While this factor is relatively low in most cases, it can nevertheless be deleterious for high-precision sensing tasks. However, despite this, one must make a clear distinction between this frequency shift and the one in Eq.~(9). The latter one scales linearly with the system's parameters showing that finite strength QEC amplifies the bias that the noise adds to the frequency.


\section{Generalization using discrete evolution of the system} \label{sec:general}
  In the main manuscript, we generalized the bias added via QEC to an arbitrary protocol and encoding using a discrete representation of an error-corrected sensing scheme. This means that the quantum state of the sensor evolves with linear completely positive trace preserving (CPTP) maps which are successively applied to it.

  Any sensing protocol could be decomposed into three quantum channels: the noise $\cl{N}$, the sensing $\cl{U}_{\delta\tau}$ and the detection-correction $\cl{C}$ channels. For simplicity, we represent the noise processes as a Pauli channel, i.e. $\cl{N}(\rho)=\sum_k c_k N_k\,\rho\,N_k^\dagger$ with $N_k$ being generalized Pauli operators (or generators of the $n$-qubit Lie algebra $\n{SU}(n)$) and $c_k$ its probability to occur. With the normalization condition of Kraus maps and $N_k$'s orthonormality, we have that $\sum_k c_k = 1$. Although this noise model is simplistic, previous works showed that any noise process can be methodically approximated by a Pauli channel without introducing new errors~\cite{wallman_noise_2016,harper_efficient_2020,ware_experimental_2021}. For our purposes, we represent this map as following
  \be \label{eq:noise_separation}
  \cl{N}=p_\cl{I}\,\cl{I}+p_\cl{N}\cl{N}' \qquad \n{with} \qquad \cl{N}'(\rho)=\sum_k c_k' N_k\,\rho\,N_k^{\dagger} \qquad \n{and} \qquad p_\cl{N} = \sum_k c_k' = 1 - p_\cl{I} \,\,.
  \ee
  Note that $p_\cl{N}$ represents the sum of all the noise processes in the system and $p_\cl{I}$ the probability that the state stays unharmed. Moreover, $\cl{N}'$ is still a Pauli channel but with modified weights, $c_k' \equiv c_k/p_\cl{N}$.

  The sensing time intervals are represented as unitary channels $\cl{U}_{\delta\tau}(\rho)=U(\delta\tau)\,\rho\, U(\delta\tau)^\dagger$ where ${U(\delta\tau)=\exp\left(-i H\,\delta\tau \right)}$ is the unitary evolution for a time $\delta\tau$ and $H$ is the system's Hamiltonian containing the couplings to the signal to sense. Finally, the detection and correction processes can be written as ${\cl{C}(\rho)=\sum_{j} U_j\, P_j\, \rho \, P_j^\dagger\,U_j^\dagger}$ with orthogonal projectors $\{P_j\}$ representing a measurement and unitaries $\{U_j\}$ characterizing the feedback.

  The evolution of the sensor's state for a sensing time $\tau$ would then be calculated by $c$ applications of the map ${\cl{C}\circ \,\cl{U}_{\delta\tau} \circ \cl{N}}$ to the initial state $\rho_0$ such that $\tau=c\,\delta\tau$. We can write it formally as a product of linear maps and simplify it further using the noise channel given in Eq.~\eqref{eq:noise_separation},
  \be \label{eq:disc_evolution}
  \begin{split}
    \rho (\tau) &= \left[\,\prod_{k=1}^{c} \cl{C} \circ \,\cl{U}_{\delta\tau} \circ \cl{N} \right]\!\bigl(\rho_0\bigr)\,\,
    =\biggl[\,p_\cl{I}\,\cl{C}\circ\,\cl{U}_{\delta\tau} + p_\cl{N}\,\cl{C} \circ \,\cl{U}_{\delta\tau} \circ \cl{N}' \biggr]^c\bigl(\rho_0\bigr) = \\
    &= \sum_{k=0}^{c} \begin{pmatrix} c \\ k \end{pmatrix} p_\cl{I}^{c-k}\,p_\cl{N}^k \biggl[\cl{C} \circ \,\cl{U}_{\delta\tau} \circ \cl{N}' \biggr]^{k} \, 
    \biggl[\cl{C} \circ \,\cl{U}_{\delta\tau} \biggr]^{c-k} \,\bigl(\rho_0\bigr) =\\
    &= \sum_{k=0}^{c} \begin{pmatrix} c \\ k \end{pmatrix} p_\cl{I}^{c-k}\,p_\cl{N}^k \biggl[\cl{C} \circ \,\cl{U}_{\delta\tau} \circ \cl{N}' \biggr]^{k} \, 
    \biggl[\,\cl{U}_{\delta\tau}^{-1} \biggr]^{k}\,\Bigl(\cl{U}_{\tau}\bigl(\rho_0\bigr)\Bigr) =\\
    &= \sum_{k=0}^{c} \begin{pmatrix} c \\ k \end{pmatrix} p_\cl{I}^{c-k}\,p_\cl{N}^k \biggl[\cl{C} \circ \,\cl{U}_{\delta\tau} \circ \cl{N}'\circ\,\cl{U}_{\delta\tau}^{-1} \biggr]^{k}\,\bigl(\rho_{\tau}\bigr) \,\,.
  \end{split}
  \ee
  This derivation relies as well on two additional assumptions. First of all, as in the continuous scenario, $\rho$ is defined in the logical subspace of the Hilbert space. Secondly, the binomial theorem for CPTP maps is permitted as long as the latter commute, which in our case means
  \be \label{eq:bin_thm_condition}
    \bigl[\,\cl{C}\circ\,\cl{U}_{\delta\tau}\,,\,\,\cl{C} \circ \,\cl{U}_{\delta\tau} \circ \cl{N}'\,\bigr](\rho)=0 \,\,.
  \ee
  This questions the relevance of the order between error-free and erroneous cycles. With an ideal detection and correction procedure, i.e. $\cl{C}\circ\cl{N}(\rho_L)=\rho_L$, Eq.~\eqref{eq:bin_thm_condition} is fulfilled for any logical density matrix $\rho_L$. If instead the noise errors are not perfectly corrected, in other words $\cl{C}\circ\cl{N}(\rho_L)=(1-\epsilon)\rho_L+\epsilon\,\cl{N}_\epsilon(\rho_L)$ for $\epsilon\in(0,1)$, then the commutation relation is likely to be unsatisfied. However, in realistic QEC experiments, we believe that higher-order errors outlined as $\cl{N}_\epsilon$ are rare ($\epsilon\ll1$) and that they can be neglected in the state evolution. A thorough study of the condition in Eq.~\eqref{eq:bin_thm_condition} is nevertheless required to support our assertion for an arbitrary encoding and sensing unitary map.

  In Eq.~\eqref{eq:disc_evolution}, we reduced $\cl{C}\circ\,\cl{U}_{\delta\tau}$ to simply $\cl{U}_{\delta\tau}$ because of the practical conjecture that the sensing Hamiltonian $H$ does not alter the encoding used in the protocol. This means that one can split it into projections onto the logical and erroneous subspace. The difference in these two parts is the origin of the bias. Mathematically, it manifests as the non-commutation of $\cl{U}_{\delta\tau}\circ\cl{N'}$ and $\cl{U}_{\delta\tau}^{-1}$. So the biasedness of an error-corrected quantum sensing protocol due to the correction steps is ultimately verified via
  \be \label{eq:bias_condition}
    \bigl[\,\cl{U}_{\delta\tau}\circ\cl{N'},\,\cl{U}_{\delta\tau}^{-1}\,\bigr](\rho_L)\neq0 \,\,.
  \ee

  Finally, we can observe that the CPTP map which is applied to the noiseless state $\rho_\tau$ in Eq.~\eqref{eq:disc_evolution} represents an expected value of a function of a random variable $X$ following a binomial distribution $\n{Bin}(c,p_\cl{N})$. This exponential function can be Taylor expanded around an arbitrary value $\beta$,
  \be \label{eq:bias_evolution}
  \begin{split}
    \rho (\tau) &= \mathbb{E}\Bigl[ \bigl(\cl{C} \circ \tilde{\cl{N}} \bigr)^{X} \Bigr] \,\bigl(\rho_{\tau}\bigr) = \sum_{j=0}^\infty\,\frac{1}{j!}\,\mathbb{E}\Bigl[ \bigl( X-\beta\bigr)^j \Bigr] \,
    \ln^{\,j}\Bigl[\cl{C} \circ \tilde{\cl{N}} \Bigr]\,
    \Bigl(\cl{C} \circ \tilde{\cl{N}} \Bigr)^{\beta}
    \,\bigl(\rho_{\tau}\bigr) =\\
    &= \Bigl(\cl{C} \circ \tilde{\cl{N}} \Bigr)^{\beta}
    \,\bigl(\rho_{\tau}\bigr) + 
    \Bigl(\mathbb{E}[ X ] - \beta \Bigr) \,\ln\Bigl[\cl{C} \circ \tilde{\cl{N}} \Bigr]\, 
    \Bigl(\cl{C} \circ \tilde{\cl{N}} \Bigr)^{\beta}\,\bigl(\rho_{\tau}\bigr) + \, ...
  \end{split}
  \ee
  with $\tilde{\cl{N}}=\cl{U}_{\delta\tau} \circ \cl{N}'\circ\,\cl{U}_{\delta\tau}^{-1}=\sum_k c_k' \tilde{N}_k\,(\cdot)\,\tilde{N}_k^{\dagger}$ where $\tilde{N}_k = U(\delta\tau) N_k U(\delta\tau)^\dagger$ are the generalized Pauli operators in the interaction picture of the system's Hamiltonian.

  A realistic value for the expansion center $\beta$ is to take the mode of the random variable $X$, since it represents the maximum of the probability density function and in other words the most frequent value of $X$. For $X\sim\n{Bin}(c,p_\cl{N})$, this means $\beta=\mathrm{Mode}\bigl[X\bigr]=\lfloor(c+1)p_\cl{N}\rfloor$ where $\lfloor\cdot\rfloor$ is the floor function. Another practical value for the expansion center is the expected value of $X$, i.e. $\beta=\mathbb{E}\bigl[X\bigr]=c\,p_\cl{N}$, because in this case one can express the evolution of the density matrix using the central moments of the random variable, $\mu_j(X)=\mathbb{E}\Bigl[ \bigl( X-\mathbb{E}\bigl[X\bigr])^j \Bigr]$. Fortunately, the mode and the mean of binomial distributions approximately coincide. We can thus conclude that
  \be \label{eq:bias_evolution_approx}
    \rho (\tau) \approx \Bigl(\cl{C} \circ \tilde{\cl{N}} \Bigr)^{c\,p_\cl{N}} \,\bigl(\rho_{\tau}\bigr)
  \ee
  up to the second order in terms of $p_\cl{N}$ (since $\mu_1(X)=0$ for $X\sim\n{Bin}(c,p_\cl{N})$\,). Eq.~\eqref{eq:bias_evolution_approx} expresses the evolution of logical states undergoing $c\,p_\cl{N}$ (among $c$) noisy but corrected cycles. The exponential decay due to uncorrected errors is carried out by the logarithmic map from Eq.~\eqref{eq:bias_evolution} which has a negative spectral radius meaning that it will produce negative terms in the infinite sum and thus deteriorate the ideal density matrix $\rho_\tau$. 

  Conclusively, to determine the amount of bias introduced by QEC, one must evaluate the difference of the expectation value of the measured logical operator $O_L$,
  \be \label{eq:bias_general}
    \Delta_{O_L} = \n{Tr}\Bigl(O_L\,\rho (\tau)\Bigr) - \n{Tr}\Bigl(O_L\,\rho_\tau\Bigr) \approx 
    \n{Tr}\Bigl(O_L\,\Bigl(\cl{C} \circ \tilde{\cl{N}} \Bigr)^{c\,p_\cl{N}}\!\bigl(\rho_{\tau}\bigr) \Bigr) - \n{Tr}\Bigl(O_L\,\rho_\tau\Bigr)\,\,.
  \ee
  The choice to measure expected values of logical operators instead of physical ones is motivated by the desire to emulate a noiseless single-qubit quantum sensing protocol. In the case of the error-corrected Ramsey scheme presented in the main text, we refer to the bias as the difference in frequencies between the ideal evolution and the actual signal. It would then be expressed as the difference of arguments of the maxima (i.e. argmax) of Fourier transforms of expectation values from Eq.~\eqref{eq:bias_general}. In the following section, we present the discrete scenario solution to our prime example and compare it to the master equation one presented in Section~\ref{sec:sol_err_corr}.

\section{Solution of the discrete scenario} \label{sec:sol_discr}
  In the previous section, we presented a general procedure to determine whether a sensing protocol will be biased by a specific QEC code or not, as well as a straightforward procedure to obtain the output density matrix deduced from the corrected errors. Let us apply it to our archetypal protocol: 

  The setting of the discrete scenario is as following
  \be \label{eq:setting_discr}
  \begin{split}
    &\cl{U}_{\delta\tau}(\rho) = U(\delta\tau)\,\rho\,U(\delta\tau)^\dagger
    \qquad \n{with} \qquad U(\delta\tau)=\exp\biggl(-i\frac{\omega}{2}\delta\tau\sum_{j=1}^{3}\sigma_z^{(j)}\biggr) 
    \qquad \n{and} \\
    &\cl{C}(\rho) = P_0\,\rho\,P_0^\dagger + \sum_{j=1}^{3} \sigma_x^{(j)}\,P_j\,\rho\,P_j^\dagger\,\sigma_x^{(j)}
    \quad \n{with} \quad P_0=\ket{0}_L\bra{0}_L + \ket{1}_L\bra{1}_L 
    \quad \n{and} \quad P_j = \sigma_x^{(j)}\,P_0\,\sigma_x^{(j)}
  \end{split}
  \ee
  are the unitary evolution and detection and correction channels, respectively. Regarding the error channel, there exist two implementations of the bit-flip noise processes
  \be \label{eq:noise_opt_real}
  \begin{split}
    &\cl{N}_{\n{opt}}(\rho) = \bigl(1 - p_\cl{N}\bigr) \rho + 
    p_\cl{N} \sum_{j=1}^{3} \frac{1}{3} \sigma_{x}^{(j)}\,\rho\,\sigma_{x}^{(j)} \qquad \n{and} \\
    &\cl{N}_\n{real}(\rho) = \bigl(1 - p_\cl{N}\bigr) \rho + \sum_{j=1}^{3} p \, \sigma_{x}^{(j)} \, \rho \,\sigma_{x}^{(j)} + 
    \sum_{k\neq l} p^2 \, \sigma_{x}^{(k)} \,\sigma_{x}^{(l)} \, \rho \, \sigma_{x}^{(l)} \, \sigma_{x}^{(k)} + 
    p^3 \, \sigma_{x}^{L} \, \rho \, \sigma_{x}^{L} \,\,.
  \end{split}
  \ee
  The former one represents an optimal Pauli channel which assumes a null probability for weight-2 and -3 errors to occur; $\cl{N}_\n{real}$ is in turn a more realistic channel since it considers arbitrary combinations of bit-flips each occuring with a probability $p$. Consequently, in the latter case, the total probability that the system gets deteriorated (cf. Eq.~\eqref{eq:noise_separation}) is $p_\cl{N}=3p+3p^2+p^3$. In the following derivations, we will exclusively consider $\cl{N}_\n{opt}$, but some simulation results using the realistic noise channel will nevertheless be discussed.

  The quantity of interest in an error-corrected Ramsey sequence as mentioned previously is the evolution of the coherence $q(\tau)=\bra{1}_L\rho(\tau)\ket{0}_L$. It is straightforward to see that the three aforementioned quantum channels $\cl{U}_{\delta\tau}$, $\cl{C}$ and $\cl{N}_{\n{opt}/\n{real}}$ satisfy the commutation relation needed for applying the binomial theorem given in Eq.~\eqref{eq:bin_thm_condition}, we can thus write it as
  \be \label{eq:q_evolution_discr}
    q(\tau) = \frac{1}{2}\,\bexp^{\,i\,3\,\omega\,\tau} \,\, \sum_{k=0}^{c} \begin{pmatrix} c \\ k \end{pmatrix} p_\cl{I}^{c-k}\,p_\cl{N}^k \,\, \bra{0}_L \biggl[\cl{C} \circ \,\cl{U}_{\delta\tau} \circ \cl{N}'_\n{opt}\circ\,\cl{U}_{\delta\tau}^{-1} \biggr]^{k} \bigl(\ket{0}_L + \bexp^{-i\,3\,\omega\,\tau} \ket{1}_L\bigr) \,\, .
  \ee
  Here we used the fact that the noiseless evolution of the density matrix is $\rho_\tau = \ket{\psi(\tau)}\bra{\psi(\tau)}$ with $\ket{\psi(\tau)}={\frac{1}{\sqrt{2}}\bigl(\ket{0}_L + \bexp^{-i3\omega\tau} \ket{1}_L\bigr)}$. Since the optimal error channel $\cl{N}_\n{opt}$ does not include weight-2 and -3 processes, we can drop the $\ket{1}_L$ element from the sum. Moreover, the superoperator $\cl{C} \circ \,\cl{U}_{\delta\tau} \circ \cl{N}'_\n{opt}\circ\,\cl{U}_{\delta\tau}^{-1}$ leaves the logical subspace invariant and while acting on $\ket{0}_L$ only induces a global phase of $\bexp^{-i\,2\,\omega\,\delta\tau}$. All that gives rise to the following result
  \be \label{eq:q_evolution_discr_bin}
    q(\tau) = \frac{1}{2}\,\bexp^{\,i\,3\,\omega\,\tau} \,\, \sum_{k=0}^{c} \begin{pmatrix} c \\ k \end{pmatrix} p_\cl{I}^{c-k}\,p_\cl{N}^k \,\bexp^{-i\,2\,\omega\,\delta\tau\,k} \equiv \frac{1}{2}\,\bexp^{\,i\,3\,\omega\,\tau} \,\, \mathbb{E}\Bigl[\bexp^{-i\,2\,\omega\,\delta\tau\,X}\Bigr] \quad \n{with} \quad X\sim\n{Bin}(c,p_\cl{N}) \,\, .
  \ee
  For a great enough number of cycles $c$ and error probabilities $p_\cl{N}$ away from 0 and 1 (some specific precision bounds can be obtained using the Berry-Esseen theorem \cite{berry_accuracy_1941,esseen_liapunov_1942}), one can effectively approximate the binomial distribution with a normal law with a mean value at $c\,p_\cl{N}$ and a variance of $c\,p_\cl{N}\,p_\cl{I}$. In that case, the expected value equals to $\exp\bigl[2\,\omega\,\tau\,p_\cl{N}(-i-p_\cl{I}\,\omega\,\delta\tau)\bigl]$ and the output of the Ramsey sequence becomes
  \be \label{eq:my_disc_sol_normal}
    \langle \sigma_x^L \rangle = 2 \n{Re}(q(\tau)) \approx \bexp^{-2[\,p_\cl{N}\,p_\cl{I}\,\omega^2\,\delta\tau]\,\tau}\,\cos\bigg[3\,\omega\Big(1 - \frac{2}{3}p_\cl{N}\Big)\tau\bigg]\,.
  \ee
  Eq.~\eqref{eq:my_disc_sol_normal} has a similar form as Eq.~(8). We can indeed identify a frequency bias and a reduced error rate. In order to make a parallel with the effective parameters given in Eq.~(9), we replace $p_\cl{N}=3p$ with $p$ being the bit-flip probability of a single physical qubit and identify $\Gerr\sim \frac{p}{\delta\tau}$ and $\Gqec\sim\frac{1}{\delta\tau}$, then it implies that $p\sim\frac{\Gerr}{\Gqec}$. The decay rate has however an extra term $p_\cl{I}(\omega\,\delta\tau)^2$ which potentially encapsulates higher-order corrections of $\Geff$ (cf. Section~\ref{sec:higher_ord_approx}),
  \be \label{eq:my_disc_sol_simplified}
    \langle \sigma_x^L \rangle = \bexp^{-3 \,\Geff\,\tau\, \, [p_\cl{I}(\omega\,\delta\tau)^2]}\,\cos[\,3\,\oeff\,\tau\,]\,.
  \ee

  \begin{figure}[t!]
    \centering
    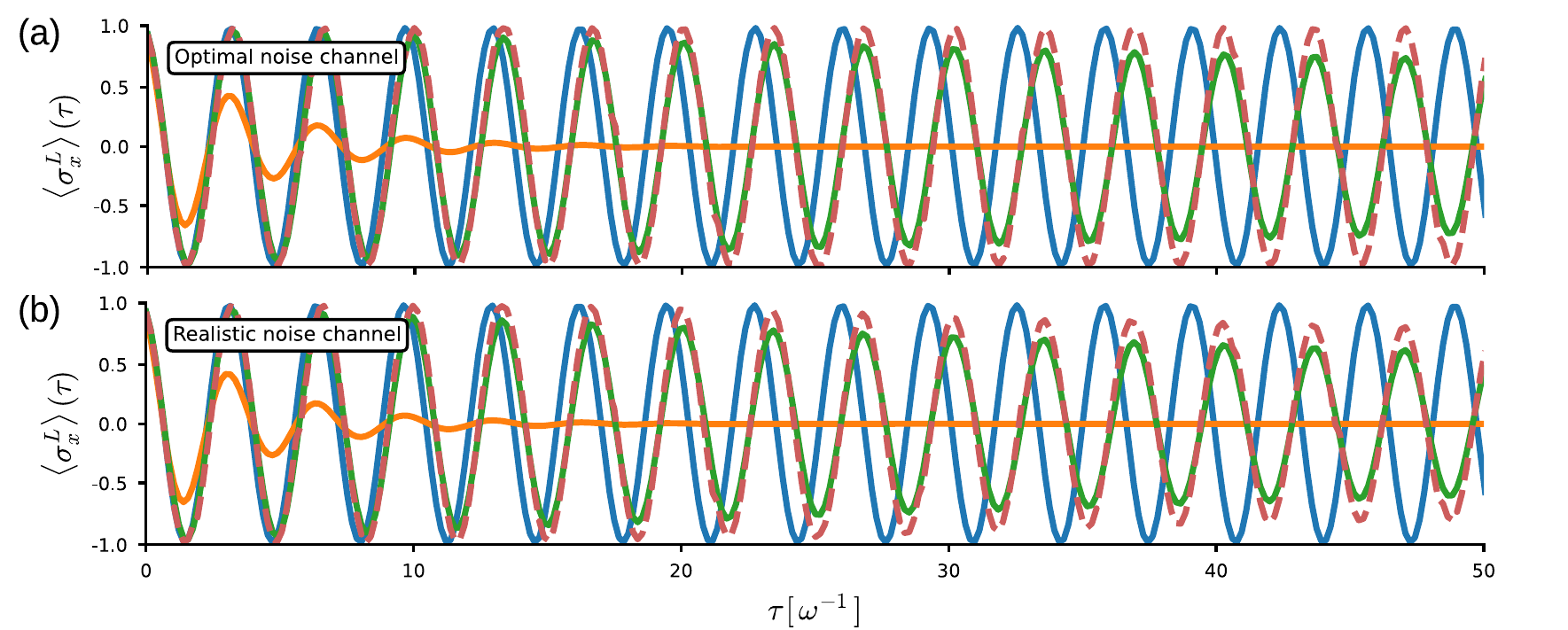
    {\phantomsubcaption\label{fig:discrete_evolution_A}}
    {\phantomsubcaption\label{fig:discrete_evolution_B}}
    \caption{ \textit{Simulation of the discrete version of the error corrected Ramsey signal.} We perform the simulation using consecutive applications of quantum channel presented in Eqs.~\eqref{eq:setting_discr} and~\eqref{eq:noise_opt_real}. (a) was obtained using the optimal noise process $\cl{N}_\mathrm{opt}$ and (b) using the realistic one $\cl{N}_\mathrm{real}$. The color scheme is identical to Fig.~1(b). The red dashed line illustrate the simulation of the dynamics subtracted from the bias given in Eq.~\eqref{eq:bias_evolution_approx}. }
    \label{fig:discrete_evolution}
  \end{figure}

  As explained in the previous section, the unbiased evolution deprived of the exponential decay is obtained using Eq.~\eqref{eq:bias_evolution_approx}. In the optimistic case considered here, it comes down to
  \be \label{eq:my_disc_sol_unbiased}
    \rho (\tau) \approx \ket{\Psi(\tau)}\bra{\Psi(\tau)}  \qquad \n{with} \qquad 
    \ket{\Psi(\tau)}=\frac{1}{\sqrt{2}}\Bigl( \ket{0}_L + \bexp^{-i\,3\,\omega\,\tau+i\,2\,\omega\,\delta\tau\,c\,p_\cl{N}} \ket{1}_L\Bigr)
  \ee
  up to a global phase. This means that the expectation value of the logical Pauli-x operator is a cosine function with a frequency $3\,\omega(1-\frac{2}{3}\,p_\cl{N})$ which is equivalent to the effective frequency as argued above.

  \figref{fig:discrete_evolution_A} shows the evolution of $\langle \sigma_x^L \rangle$ in the discrete scenario of the error-corrected Ramsey protocol. On top of the usual three situations, viz. the ideal, erroneous and error-corrected (the color code is preserved), it also exhibits the unbiased evolution of the system obtained using Eq.~\eqref{eq:bias_evolution_approx} where for simplicity we rounded the power to the closest integer. We can indeed observe that the latter signal oscillates with the same frequency as the green curve without decaying. \figref{fig:discrete_evolution_B} depicts similar situation but simulated with the realistic noise channel $\cl{N}_\n{opt}$. We can see that in this case the unbiased evolution still captures the effective frequency. However, it is exponentially decaying in amplitude due to the presence of weight-2 and -3 processes which are left uncorrected. They essentially break the commutation relation given in Eq.~\eqref{eq:bin_thm_condition} due to the fact that $\cl{C}\circ\cl{N}_\mathrm{real}(\rho_L)=(1-\epsilon)\rho_L+\epsilon\,\cl{N}_\epsilon(\rho_L)$ with $\epsilon=3p^2+p^3$ and $\cl{N}_\epsilon (\rho_L) = \sigma_{x}^{L} \, \rho \, \sigma_{x}^{L}$. Nevertheless, we can observe that some aspects of the analysis performed in Section~\ref{sec:general} can still be useful in some cases which points out that it can be expanded to a broader range of situations.
\section{Example of an unbiased sensing protocol} \label{sec:gkp_sensing}
The generalization of the QEC-induced bias raises the question of the existence of unbiased sensing protocols. Eq.~\eqref{eq:bias_condition} sets a condition that an encoding should satisfy in order to produce an accurate output signal. Solving this commutation relation would give rise to a general procedure for finding such codes, but that goes beyond the scope of this work. We can however provide an example of an error-corrected sensing protocol that would induce no bias after correction steps.

In the past few year, one has observed great experimental advances in preparation, manipulation and correction of logical states encoded in harmonic oscillator states \cite{fluhmann_sequential_2018,fluhmann_encoding_2019,campagne-ibarcq_quantum_2020,de_neeve_error_2020}. One of the most prominent platforms are trapped ions where qubits are encoded in different motional modes of the charged atom. We can thus imagine to sense small amplitude electric or magnetic fields as well as mechanical forces which can be effectively expressed as displacements in the phase space of the harmonic oscillator. In fact, some sensors based on motional modes of ions were already proposed in the recent literature \cite{maiwald_stylus_2009,campbell_rotation_2017,biercuk_ultrasensitive_2010,shaniv_quantum_2017,duivenvoorden_single-mode_2017,wolf_motional_2019}.

In the case of an electric field, it couples to the harmonic oscillator with the following Hamiltonian~\cite{fluhmann_thesis_2019}
\be \label{eq:E_field_Ham}
  H \approx \frac{i\,g}{2} \Bigl( \bexp^{i \, \Delta\omega\, t + i\,\varphi } \, a + \bexp^{-i \, \Delta\omega\, t - i\,\varphi } \, a^\dagger \Bigr)
\ee
with $a$ and $a^\dagger$ being the creation and annihilation operators, $\varphi$ the phase of the electric field, $\Delta\omega$ the frequency detuning of the field from the motional mode and $g$ their coupling strength. The latter quantity is set by the amplitude of the electric signal and the trap geometry. We would then be able to sense every parameter of the field (i.e. amplitude, frequency and phase). On resonance, after integrating over a time window $\delta\tau$, the evolution is described in the phase space as a displacement
\be \label{eq:E_field_displ}
  U(\delta\tau) = D\bigl(\alpha(\delta\tau)\bigr) \qquad \n{with} \qquad D(\alpha) = \bexp^{\alpha\,a^\dagger-\alpha^*\,a} \qquad \n{and} \qquad \alpha(\delta\tau) = -\frac{i}{2} \,g\,\delta\tau\,\bexp^{i\,\varphi} \,\,.
\ee
Let us for the remainder of this Section assume that one would like to sense for the phase $\varphi$ of the electric field only. We are thus interested in determining the direction of the displacement.

A suitable encoding to detect and correct displacements is the so-called Gottesman-Preskill-Kitaev (GKP) code~\cite{gottesman_encoding_2001}. Without entering in the details of the encoding itself, it is represented as a superposition of Gaussian states arranged in a grid-like style in the phase space and allows to correct for small shifts in the $q \propto a+a^\dagger$ and $p \propto a-a^\dagger$ quadratures. However, correcting for both quadratures would destroy the information about the electric field's phase that one would like to sense, we are thus bound to correct only for shifts in $q$ or $p$. 

\begin{SCfigure}[][t!]
  \centering
  \includegraphics[width=.34\linewidth]{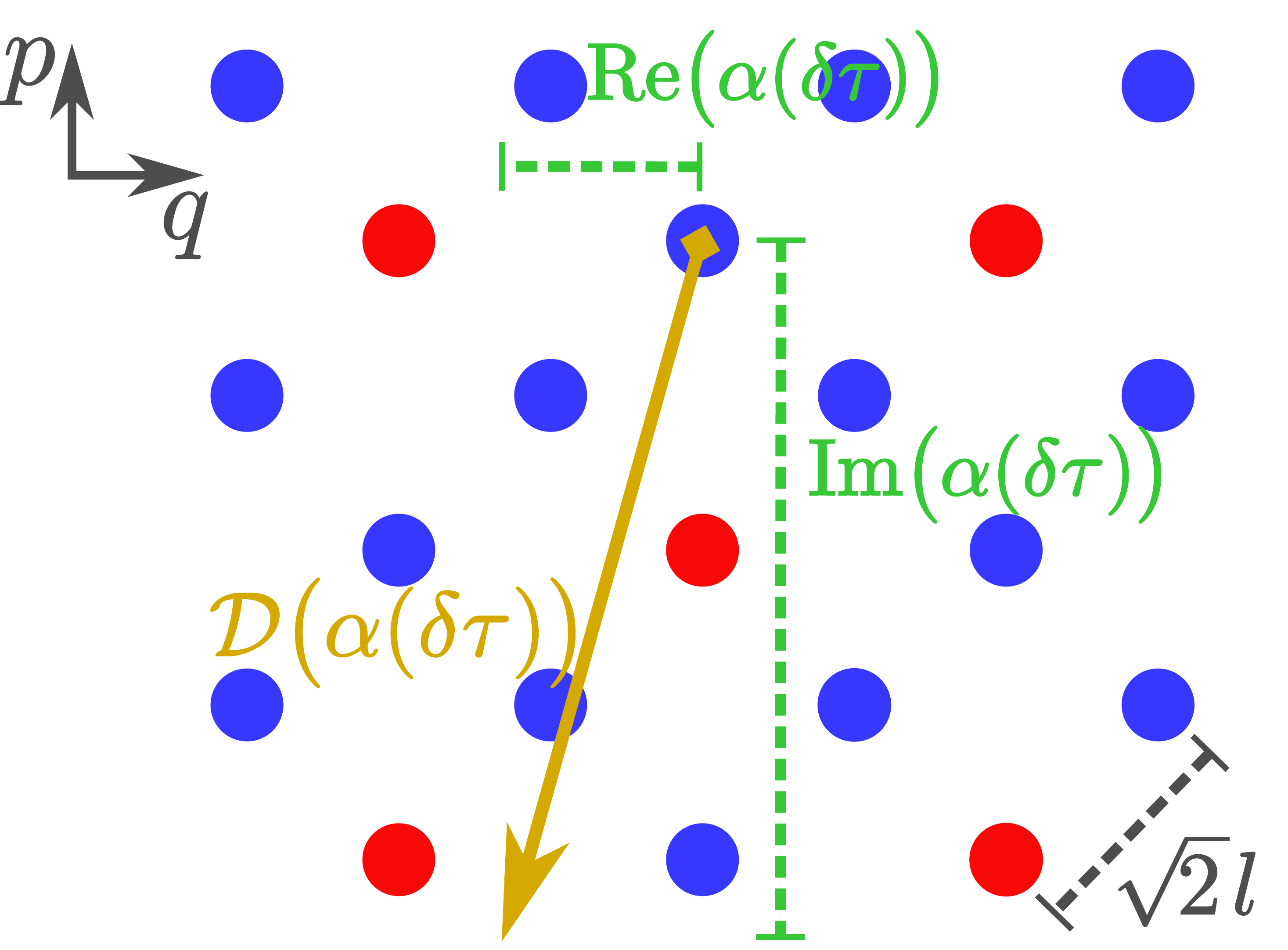}
  \caption{\textit{Sensing using GKP encoding.} Phase space representation of a grid state (blue and red dots) of latice size $l$ subjected to a displacement $D\bigl(\alpha(\delta\tau)\bigr)$ caused by an electric field (cf. Eqs.~\eqref{eq:E_field_Ham} and~\eqref{eq:E_field_displ}). It is decomposed into its real and imaginary parts which are related to the shifts that the state encounters while being corrected in q or p quadratures, Eq.~\eqref{eq:Delta_displ}.}
  \label{fig:gkp_sensing}
\end{SCfigure}

The protocol will then resemble the Ramsey sequence for the three-qubit repetition code illustrated in Fig.~1(a), except that we prepare the system in an eigenstate of the logical Pauli-y operator which ensures that the correction process does not erase the information about the signal if it is parallel to the corrected quadrature. Interspersing electric field sensing moments with q-correction (p-correction) processes will at every cycle displace the state in the conjugate p (q) direction by (cf.~\figref{fig:gkp_sensing}) 
\be \label{eq:Delta_displ}
  \Delta_q = \frac{\n{Im}\bigl(\alpha(\delta\tau)\bigr)}{\sqrt{2}\,l} \qquad 
  \left(\Delta_p = \frac{\n{Re}\bigl(\alpha(\delta\tau)\bigr)}{\sqrt{2}\,l} \right) \,\, .
\ee
Here $l$ is the lattice constant. Dividing the value $\Delta_q$ ($\Delta_p$) by the sensing time interval $\delta\tau$, we obtain the frequency $\mathit{f}_q$ ($\mathit{f}_p$) with which the system will oscillate between the two eigenstates of the logical Pauli-y operator, i.e. the frequency of the expectation value $\langle \sigma_Y^L \rangle$. The two frequencies $\mathit{f}_q$ and $\mathit{f}_p$ allow us to fully determine the phase $\varphi$ of the electric field.

The main sources of noise in such systems are excitation losses, heating and dephasing processes as well as random diffusive displacements~\cite{gottesman_encoding_2001}. The particularity is that they can all be expressed in terms of small random displacements which effectively contribute to increase the variance of the grid points. Since both the signal and noise processes act uniformly in the whole phase space, the logical (i.e. the grid) and erroneous (i.e. the interspace between points of the grid) subspaces evolve identically, the correction steps will not alter the output of the protocol. Nonetheless, a formal proof of their commutation as stated in Eq.~\eqref{eq:bias_condition} is required to fully support this claim.

\section{Cram\'{e}r-Rao bounds} \label{sec:cramer_rao}
  
  As mentioned in the main text, the Cram\'{e}r-Rao bound (CRB) constitutes a lower bound for the variance of an estimator~\cite{cramer_mathematical_1999}. In a univariate estimation problem, i.e. the estimator is a scalar, this bound looks, for a completely arbitrary estimator $\hat\theta$, as following:
  \be \label{eq:cramer_rao_general}
		\n{Var}(\hat\theta) \geq \frac{(1+\frac{\n{d}b}{\n{d}\theta})}{N\,\cl{I}(\theta)} \qquad\textnormal{with}\qquad 
		\cl{I}(\theta) \coloneqq \mathbb{E}\left[\left(\frac{\partial}{\partial\theta}\log\proba(X|\theta)\right)^2\right]
	\ee
  where $\{X_i\}_{i=1}^{N}$ is a set of independent and identically distributed observations drawn from a probability density function $\proba(X_i|\theta)$;\,\, $\cl{I}(\theta)$ is known as the Fisher information (FI) and describes a measure of the amount of information that the set $\{X_i\}_{i=1}^{N}$ embody about $\theta$. As a reminder, the estimator is defined as $\mathbb{E}[\hat\theta] = \theta + b(\theta)$ with $b(\theta)$ -- its bias; and its variance is given by $\n{Var}(\hat\theta) = \mathbb{E}[(\hat\theta - \mathbb{E}[\hat\theta])^2]$. In the case of an unbiased estimator, the CRB will end up to be simply $(N\cl{I}(\theta))^{-1}$.
  
  However, as indicated in the paper, the problem we are considering is multivariate since the expectation value depends both on $\Gerr$ and $\omega$ which are a priori unknown parameters. The estimator given in Eq.~(4) becomes then
  \be \label{eq:estimator_multi}
  \hat\Theta =
  \begin{pmatrix}
      \,\hat\theta_1\, \\ \,\hat\theta_2\,
  \end{pmatrix}=\underset{\{\theta_1\,,\,\theta_2\}}{\arg\min} 
  \sum_\tau \Big[\,X_\tau - \langle \sigma_x^L (\tau,\theta_1,\theta_2,\Gqec) \rangle\,\Big]^2 
  \ee
  where the components of the estimator $\hat\Theta$ can be respectively identified as the estimator of the frequency $\oest$ and of the error rate $\Gest$. Then, in this case, one can rewrite the CRB as follows:
  \be \label{eq:cramer_rao_multi}
			\n{\mathbf Cov}(\hat\Theta) - \frac{1}{N}\,\n{\mathbf I}(\Theta)^{-1} \geq 0
	\ee
  where now the variance is replaced by the covariance matrix $\n{\mathbf Cov}(\hat\Theta)$ and the FI by the Fischer information matrix~$\n{\mathbf I}(\Theta)$. Due to the fact that the latter is positive semi-definite, the following corollary is directly derived from Eq.~\eqref{eq:cramer_rao_multi}~\cite{liu_quantum_2019}
  \be \label{eq:cramer_rao_trace}
    \n{Tr}\Big(\n{\mathbf Cov}(\hat\Theta)\Big) \,=\, 
    \n{Var}(\hat\theta_1) + \n{Var}(\hat\theta_2) 
    \,\,\,\geq\,\,\,
    \frac{1}{N}\,\n{Tr}\Big(\n{\mathbf I}(\Theta)^{-1}\Big) 
    \,\,\,\geq\,\,\,
    \frac{1}{N}\big(\,\cl{I}(\theta_1)^{-1} + \cl{I}(\theta_2)^{-1}\,\big)
  \ee
  In this expression, $\n{Tr}(\cdot)$ denote the trace of a matrix.
  
  \begin{SCfigure}[][b!]
    \centering
    \includegraphics[width=.49\linewidth]{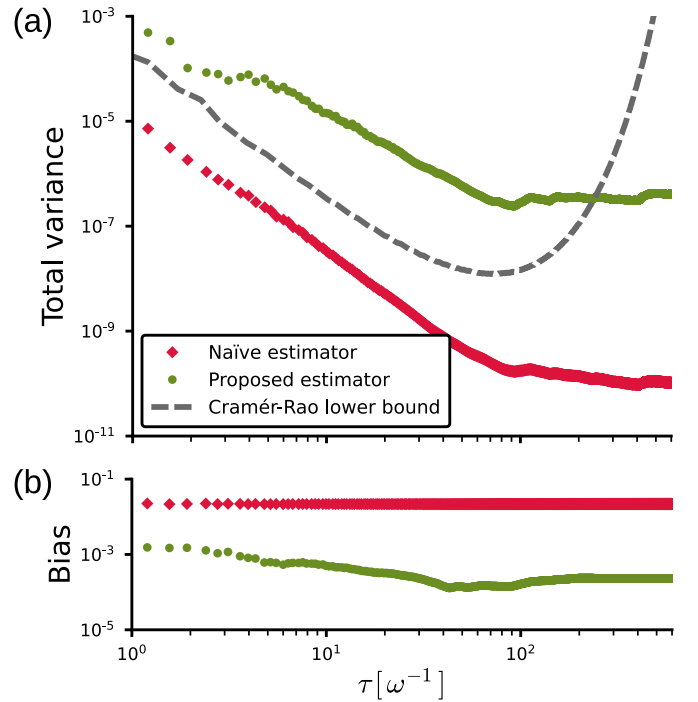}
    {\phantomsubcaption\label{fig:violation_crb}}
    {\phantomsubcaption\label{fig:bias}}
    \caption{\textit{Verification of the bias.} (a) Total variance as a function of the sensing time $\tau$ for estimators derived from Eq.~(5) (diamonds) and Eq.~(6) (circles). The right-hand side of the Cram\'{e}r-Rao lower bound given in Eq.~\eqref{eq:cramer_rao_ramsey} is presented as a dashed line. (b) Statistical bias of the frequency estimate $\hat\omega$ as a function of $\tau$. Each point from these plots was obtained from one multivariate least-squares regression of data simulated with a Monte-Carlo method with parameters $\Gerr=0.2\omega$, $\Gqec=16.6\omega$.}
    \label{fig:Fig2}
  \end{SCfigure}

  Let us derive an expression for the FI present in the CRB inequality.
  \[
		\begin{split}
			\cl{I}(\theta) &= \sum_{x=\pm1} \proba(X=x|\theta) 
			\left[\frac{\partial}{\partial\theta}\log\proba(X=x|\theta)\right]^2 = \\
			&= \proba(X=+1|\theta) \left[\frac{\partial}{\partial\theta}\log\proba(X=+1|\theta)\right]^2 + (1-\proba(X=+1|\theta)) 
			\left[\frac{\partial}{\partial\theta}\log\left(1-\proba(X=+1|\theta)\right)\right]^2 = \\
			&= \frac{1}{\proba(X=+1|\theta)} \left[\frac{\partial}{\partial\theta}\proba(X=+1|\theta)\right]^2 + 
			\frac{1}{1-\proba(X=+1|\theta)} \left[\frac{\partial}{\partial\theta}\proba(X=+1|\theta)\right]^2 = \\
			&= \frac{1}{\proba(X=+1|\theta)(1-\proba(X=+1|\theta))} \left[\frac{\partial}{\partial\theta}\proba(X=+1|\theta)\right]^2
		\end{split}
  \]
  Here, we restricted ourself specifically to our problem where the random variable is now $X$ and it can take only two values $x\in\{-1\,,\,1\}$. Moreover, the probability density function evaluated in $+1$ is straightforwardly derived from the expectation value, $\proba(X=+1|\theta) = \half \left(\langle \sigma_x^L \rangle + 1\right)$. Then equation~\eqref{eq:cramer_rao_trace} becomes
  \be
    \sum_{i=1}^2 \, \n{Var}(\hat\theta_i) 
    \,\,\,\geq\,\,\,
    \frac{1}{N} \,\left(1 + \langle \sigma_x^L \rangle\right) \left(1 - \langle \sigma_x^L \rangle\right) 
    \sum_{i=1}^2 \, \left[\frac{\partial}{\partial\theta_i} \langle \sigma_x^L \rangle\right]^{-2}.
  \ee
  In our situation, this inequality is then expressed as follows:
  \be \label{eq:cramer_rao_ramsey}
    \n{Var}(\oest) + \n{Var}(\Gest) \geq \frac{1}{N}\left(\,\cl{I}(\omega)^{-1} + \cl{I}(\Gerr)^{-1}\right) \qquad \n{with} \qquad
    \cl{I}(\lambda)^{-1} = \left(1 + \langle \sigma_x^L \rangle\right) \left(1 - \langle \sigma_x^L \rangle\right)
    \left[ \frac{\partial \langle \sigma_x^L \rangle}{\partial \lambda} \right]^{-2}\,\,.
  \ee
  where the expectation value $\langle \sigma_x^L \rangle$ is either replaced by Eqs.~(5) or (6) depending on the considered situation.

  \figref{fig:violation_crb} shows the total variance for estimators arising from both equations, obtained through fitting Monte Carlo simulations, as well as the Cram\'{e}r-Rao lower bound from Eq.~\eqref{eq:cramer_rao_ramsey}. Notice that the estimator derived from the conjectured Eq.~(5) violates this fundamental bound of estimation theory for all $\tau$ and is therefore biased. The estimator from our proposed solution in Eq.~(6) only violates the CRB after long times $\tau$ when the sensor has fully decohered (the fitting algorithm starts to fit statistical fluctuations around 0). The frequency bias can as well be statistically assessed using ${b(\omega)=\mathbb{E}[(\omega - \oest)^2] - \n{Var}(\oest)}$ that characterizes the mean square deviation of the estimate $\oest$ from the true value of $\omega$. This quantity is represented as a function of $\tau$ in \figref{fig:bias}. From here we see that the estimator using Eq.~(6) has a lower bias than the one using Eq.~(5). We thereby establish that the proposed estimator is in fact less biased that the na\"ive one which is statistically inconsistent. Additionally, we remark that it has a greater total variance but a lower bias than the estimator derived using the conjectured solution. This contrast between \figref{fig:violation_crb} and \figref{fig:bias} reflects the so-called bias-variance tradeoff, which is a generic statistical observation~\cite{geman_neural_1992}.


\section{Minimum detectable signal} \label{sec:sensitivity}
  
  \citet{degen_quantum_2017} define the minimum detectable signal as the smallest variation of $\omega$ giving a unit signal to noise ratio: 
  $\n{SNR} = \left|\delta\proba(X_i)\right| / \sigma$ where $\delta\proba(X_i)$ is the variation of the probability density function introduced in the previous section and $\sigma$ its standard deviation. Knowing that we deal with a binomial distribution and that $\delta\proba(X_i) = \delta\omega\,\partial_\omega\proba(X_i)$,
  it implies that  
  \be \label{eq:sensitivity_general}
    |\delta\omega|=\sqrt{\frac{1}{N}\,\proba(X=+1)\left[1-\proba(X=+1)\right]}\,
    \left|\frac{\partial\,\proba(X=+1)}{\partial\,\omega}\right|^{-1} \equiv \frac{1}{\sqrt{N \mathcal{I}(\omega)}}
  \ee
  where $N$ is, as before, the number of observations. The derivative of the na\"ive Eq.~(5) with respect to $\omega$ is straightforward and its $|\delta\omega|$ is thus equal to Eq.~(13). For the proposed solution given in Eq.~(6), this derivative is however more complicated due to the form of $q$. It is given by the following expression
	\be \label{eq:my_sensitivity_full}
		\frac{\partial \langle \sigma_x^L \rangle}{\partial\omega} = 2\n{Re}\big(-2i\tau\,q(\tau) + \dif q(\tau)\,\big)	
	\ee
	\[
		\begin{split}
			&\text{with \quad\quad} \dif q(\tau) = 
				\dif C_+ \,\bexp^{\,\half\tau \left(-\Gamma_\n{qec} - 6\Gamma_\n{err} - 4 i \omega + \sqrt{D}\right)} -
				\dif C_- \,\bexp^{\,\half\tau \left(-\Gamma_\n{qec} - 6 \Gamma_\n{err} - 4 i \omega - \sqrt{D}\right)} \\
			&\text{and \quad\quad} \dif C_\pm = \frac{1}{4\, D^{3/2}} \left[ -2i(D-\Gqec^{\,2}) - 
				\tau \sqrt{D}(i\Gqec+2\omega)(\pm\Gqec \mp 2i\omega + \sqrt{D})\right] \,\,.
		\end{split}
	\]
  This function can then be inserted in equation \eqref{eq:sensitivity_general} to obtain its minimum detectable signal, this step has not been done analytically due to the complexity of this expression.
  
  As an alternative, we can also derive the frequency variation $|\delta\omega|$ using the reduced solution presented in Eq.~(8). Due to the requirement of performing a partial derivative with respect to $\omega$, we must opt for higher order approximations of the effective decay rate and frequency (cf. Section~\ref{sec:higher_ord_approx}). That is to say that $\Geff$ and $\oeff$ have to be considered as function of $\omega$. Thus, the minimum frequency variation that one can sense would be given by
  \be \label{eq:my_sensitivity_reduced}
    |\delta\omega|(\tau)=
    \sqrt{\frac{1-\bexp^{-6\,\Geff\,\tau}\,\cos^2[\,3\,\oeff\,\tau\,]}
    {N\,9\,\tau^2\,\bexp^{-6\,\Geff\,\tau}\,
    \big((\partial_\omega\Geff)\,\cos[\,3\,\oeff\,\tau\,] 
    + (\partial_\omega\oeff)\,\sin[\,3\,\oeff\,\tau\,]\big)^2}}
  \ee
  An important aspect to mention is that this formula has almost the same form as the na\"ive expression in Eq.~(13), except that now we have some $\partial_\omega\oeff$ and $\partial_\omega\Geff$ terms in the denominator. A numerical study of the ratio of this terms shows that, in the validity range~(cf. Section~\ref{sec:reduced_sol}), the former extensively dominates over the second one. We then conclude that that the minimum detectable variation $|\delta\omega|(\tau)$ can be approximated using the expression presented in Eq.~(15) in the main text.

  Equations (13) and (15) explicitly show how equating $\omega$ and $\oeff$ can give misleadingly optimistic figures of merit for error-corrected quantum sensors. Indeed, if we consider the notation introduced in the previous section, then
  \be \label{eq:sensitivity_comparison}
      |\delta\omega|_\n{conj} (\tau,\hat\theta_1,\hat\theta_2) 
      \approx \left|\partial_\omega\,\oeff\right|\,|\delta\omega|_\n{mod}(\tau,\hat\theta_1,\hat\theta_2) 
  \ee
  where the unadjusted and the proposed variations are given respectively by Eq.~(13) and Eq.~(15) from the main text. This highlights that the minimum detectable signal with a given protocol is potentially overestimated, i.e. better than the true one, since the partial derivative of $\oeff$ with respect to the frequency is less than one.

  \section{Proposed solution \texttt{vs} Simulation} \label{sec:disparity}

\begin{figure}[b!]
    \centering
    \def\svgwidth{.82\linewidth}
    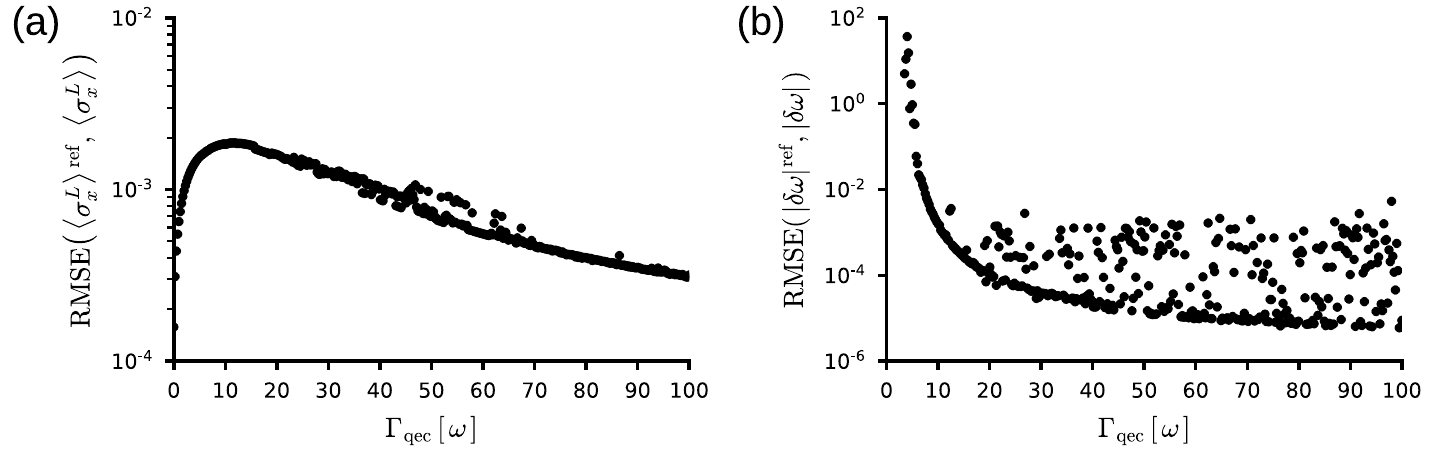
    {\phantomsubcaption\label{fig:abs_simul_proposed_A}}
    {\phantomsubcaption\label{fig:abs_simul_proposed_B}}
    \caption{ \textit{Error between the simulation and the proposed functions.} The root-mean-square error (RMSE) between the analytically derived expressions and their simulations obtained with QuTiP~\cite{johansson_qutip_2013}. These expressions are (a) the proposed function for $\langle \sigma_x^L \rangle$ (cf. Eq.~(6) in the main text), (b) the frequency variation $|\delta\omega|$ obtained with~\eqref{eq:my_sensitivity_full}. The error rate is fixed at $\Gerr=0.1\omega$.}
    \label{fig:abs_simul_proposed}
  \end{figure}

  An aspect not discussed in the main text is the precision of the proposed solution with respect to a simulation of the protocol. \figref{fig:abs_simul_proposed_A} shows the root-mean-square deviation (RMSE) of the proposed function given in Eq.~(6) from the simulated with QuTiP~\cite{johansson_qutip_2013} function. We note that the error is consistently below 0.2\% indicating that the function is an equally accurate solution of the non-corrected and error-corrected problems. We believe that this error is mainly the result of removing the orange terms from the matrix differential equation~\eqref{eq:simple_system}. 
  
  Furthermore, \figref{fig:abs_simul_proposed} displays as well the RMSE of the proposed minimum detectable signal formula presented in Eq.~\eqref{eq:my_sensitivity_full} with respect to its simulation. Notwithstanding some discrepancies for high correction rates, the baseline of the error curve depict a very fast decrease and crosses the cap of 1\% already for low $\Gqec$. The variations mentioned previously as well as the trend of the error curve when $\Gqec\rightarrow0$ could come from discretization errors. Indeed, for the simulated data, we approximated the derivative $\partial_\omega\langle \sigma_x^L \rangle$ with a finite difference method such that these errors could potentially be reduced with a finer frequency discretization or a higher order method.
  
  \section{Convergence of the reduced solution} \label{sec:higher_ord_approx}

\begin{figure}[t!]
    \centering
    \def\svgwidth{.82\linewidth}
    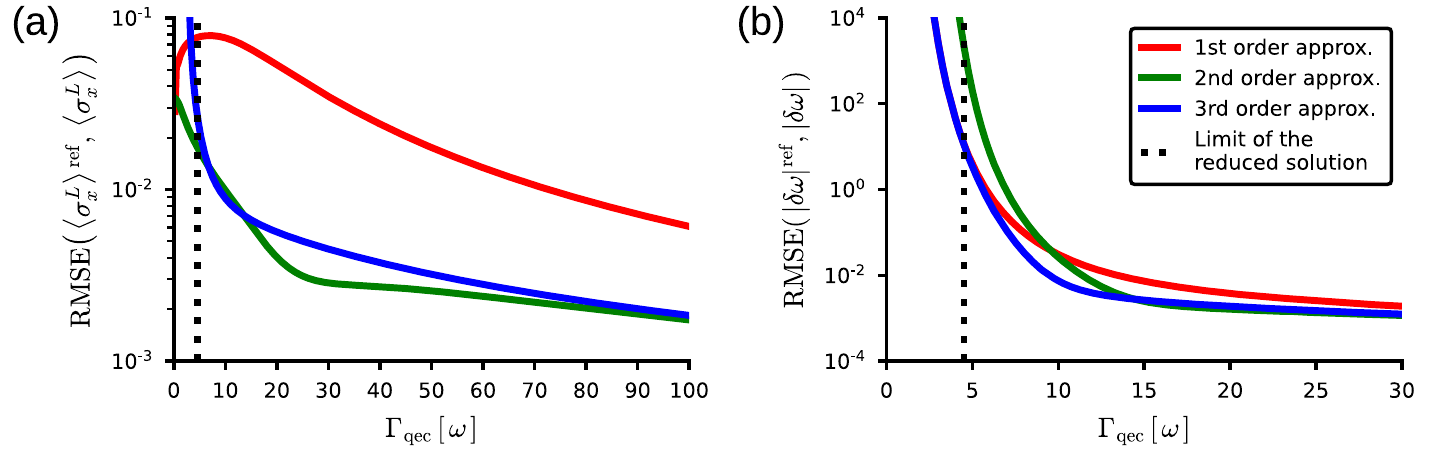
    {\phantomsubcaption\label{fig:rmse_proposed_reduced_A}}
    {\phantomsubcaption\label{fig:rmse_proposed_reduced_B}}
    \caption{ \textit{Error between the proposed solution and its approximation.} The root-mean-square error (RMSE) between: (a) the proposed function for the expectation value $\langle \sigma_x^L \rangle$ and its reduced form (respectively Eq.~(6) and (8) in the main text); (b) the frequency variation Eq.~\eqref{eq:my_sensitivity_full} and its approximation Eq.~(15). The curves differ in the order of approximation of the effective parameters as presented in Eq.~\eqref{eq:effective_param_refined}. The error rate is fixed at $\Gerr=0.1\omega$.}
    \label{fig:rmse_proposed_reduced}
  \end{figure}

  In \textit{Imperfect error-corrected sensing} part of the main text, we approximate the full expression of an error-corrected Ramsey experiment given in Eq.~(6) with a much simpler and closer to the expected form function. The latter, presented in~Eq.~(8), was derived by expanding the square root of the discriminant $D$ (cf. Eq.~\eqref{eq:expansion}) and considering only the first order terms in $\cl{O}(1/\Gqec)$. This implies that the effective parameters, stated in Eq.~(9), are as well first-order approximations of the true decay rate and frequency of the signal. This can be improved by refining the expression of $\oeff$ and $\Geff$. The next two approximation orders are the following:
  \be \label{eq:effective_param_refined}
      \begin{split}
      \omega_\n{eff} &= \omega\,\Bigg(1\,
      \underbrace{-\,2\,\frac{\Gerr}{\Gqec^{\,}}}_\textnormal{1st order}\,
      \underbrace{+\,16\,\frac{\Gerr^2}{\Gqec^2}}_\textnormal{2nd order}\,
      \underbrace{-\,141\,\frac{\Gerr^3}{\Gqec^3} +\,8\,\frac{\Gerr\,\omega^2}{\Gqec^3}}_\textnormal{3rd order}
      \Bigg)
      \\[7pt]
      \Geff &= 
      \underbrace{2\,\frac{\Gerr}{\Gqec^{\,}}\Gerr}_\textnormal{1st order}\,
      \underbrace{-\,12\,\frac{\Gerr^{2}}{\Gqec^{2}}\Gerr +\,4\,\frac{\omega^{2}}{\Gqec^{2}}\Gerr}_\textnormal{2nd order}
      \underbrace{+\,84\,\frac{\Gerr^{3}}{\Gqec^{3}}\Gerr -\,68\,\frac{\Gerr\,\omega^{2}}{\Gqec^{3}}\Gerr}_\textnormal{3rd order}
      \end{split}
  \ee
  We see that the effective decay rate now has a dependence on the true frequency $\omega$ as well. To study the precision of these approximations, one can measure the root-mean-square deviation of the resulting expectation values $\langle \sigma_x^L \rangle$ with respect to the proposed one \footnote{We do not compare these functions to their simulated counterparts as in Section~\ref{sec:disparity}, because their calculation is resource demanding.}. \figref{fig:rmse_proposed_reduced_A} illustrates this quantity as a function of the correction rate $\Gqec$. The plot shows that the error between both function is, in the scope of the validity range, below 10\% and for the 2nd and 3rd order approximations it even becomes very quickly of the magnitude of 0.1\%. The same analysis can be done on variations $|\delta\omega|$ as depicted in \figref{fig:rmse_proposed_reduced_B}. Here, the approximated functions, calculated with Eq.~(15), were compared to the full expression given in Eq.~\eqref{eq:my_sensitivity_full}. Unlike in the previous case, for low $\Gqec$, the minimum detectable signal is better approximated by the 1st and 3rd order expressions of the effective parameters than by the 2nd order one. All of them nevertheless converge relatively quickly to a same value below 1\%.

\begin{SCfigure}[][t!]
    \centering
    \def\svgwidth{.4\linewidth}
    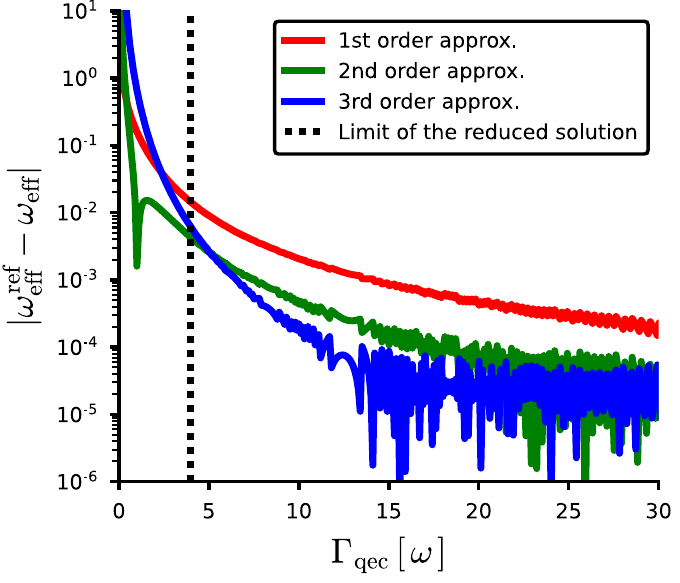
    \caption{\textit{Error of the effective frequency.} Absolute deviation of the approximated effective frequency stated in Eq.~\eqref{eq:effective_param_refined} from its true value, obtained using a discrete Fourier transform of the proposed function for $\langle \sigma_x^L \rangle$. The error rate is fixed at $\Gerr=0.1\omega$.}
    \label{fig:bias_higer_order}
  \end{SCfigure}

  The last observation suggests that the high error of the 1st order approximation in \figref{fig:rmse_proposed_reduced_A} originates from a poor approximation of the exponential decay rather than the oscillations' frequency. We can confirm this hypothesis by investigating the deviation of $\oeff$ as stated in Eq.~\eqref{eq:effective_param_refined} from the true effective frequency obtained using a discrete Fourier transform of the proposed function. \figref{fig:bias_higer_order} shows how the absolute value of this quantity changes with $\Gqec$. Since for all orders the error is below $\sim\!$1\% in the validity range and converges quickly to lower values, it constitutes an evidence that the major part of the error comes from a loose approximation of~$\Geff$.

\newpage



\end{document}